\RequirePackage[2020-02-28]{latexrelease}

\documentclass{rQUF2e}

\usepackage{epstopdf}
\usepackage{subfigure}

\theoremstyle{plain}

\theoremstyle{definition}

\theoremstyle{remark}

\usepackage{tabularx}
\newcolumntype{Y}{>{\raggedleft\arraybackslash}X}

\usepackage{booktabs}
\usepackage{makecell}
\usepackage{rotating}
\usepackage{setspace}
\usepackage{multirow}
\usepackage{csquotes}
\usepackage[colorlinks=true, allcolors=blue]{hyperref}
\usepackage{cleveref}
\usepackage[dvipsnames]{xcolor}

\definecolor{colour1}{RGB}{166,206,227}
\definecolor{colour2}{RGB}{31,120,180}
\definecolor{colour3}{RGB}{178,55,250} 
\definecolor{colour4}{RGB}{51,160,44}

\newcounter{noteMCctr} 

\usepackage{booktabs}

\newcommand{\bftab}{\fontseries{b}\selectfont}

\newcommand{\update}[1]{{\color{black} #1}}
\newcommand{\updateSecondRound}[1]{{\color{black} #1}}

\usepackage{titlesec}

\definecolor{colour3}{RGB}{178,55,250} 
\newcommand{\mc}[1]{\textcolor{black}{#1}}

\begin{document}


\title{Trade Co-occurrence, Trade Flow Decomposition, and Conditional Order Imbalance in Equity Markets}

\author{Yutong Lu$^{\ast \dag}$\thanks{$^\ast$Corresponding author.
Email: yutong.lu@stats.ox.ac.uk}, Gesine Reinert$^{\dag \odot}$
and Mihai Cucuringu$^{\dag \ddag \mathsection \odot}$}

\affil{$\dag$ Department of Statistics, University of Oxford, 24-29 St Giles', Oxford OX1 3LB, UK\\
$\ddag$ Mathematical Institute, University of Oxford, Woodstock Rd, Oxford OX2 6GG, UK \\  $\mathsection$ Oxford-Man Institute of Quantitative Finance, University of Oxford, UK  \\
$\odot$ The Alan Turing Institute, 96 Euston Rd, London NW1 2DB, UK
}


\maketitle

\begin{abstract}
The time proximity of high-frequency trades can contain a salient signal. In this paper, we propose a method to classify every trade, based on its proximity with other trades in the market within a short period of time, into five types. By means of a suitably defined normalized order imbalance associated to each type of trade, which we denote as \textit{conditional order imbalance} (COI), we investigate the price impact of the decomposed trade flows. Our empirical findings indicate strong positive correlations between contemporaneous returns and COIs. In terms of predictability, we document that associations with future returns are positive for COIs of trades which are isolated from trades of stocks other than themselves, and negative otherwise. Furthermore, trading strategies which we develop using COIs achieve conspicuous returns and Sharpe ratios, in an extensive experimental setup on a universe of 457 stocks using daily data for a period of four years. \\
\end{abstract}


\begin{classcode}C0, C2\end{classcode}

\section{Introduction}
The transformation of major equity exchanges to electronic trading significantly reshapes the market microstructure landscape, by reducing latency up to nanoseconds \citep{o2015high, hirschey2021high}, and thus leading to market participants achieving unprecedented levels of profitability in their trading strategies. Every agent in the market can directly submit and cancel limit orders. Trades are settled when existing limit orders are executed by market orders/marketable limit orders. Trades, carrying distinct information and having their own impact  on the price changes of the underlying stocks, have been classified into different types and studied separately by academics and practitioners. For example, grouping by directions of trading, \citet{chordia2016buyers} study flows of buyer- and seller-initiated trades, thus decomposing into aggressive buys and aggressive sells. \citet{kraus1972parallel} and \citet{lee2004order} separate institutional trades from trades placed by individual investors. Different from these classifications, which are exclusively based on the characteristics of the individual trades, in this paper, we classify trades according to their time of placement relative to the arrival time of other trades across the market, both within the same asset and also cross-sectionally across the available universe of stocks. \updateSecondRound{We find that the time proximity of trade arrivals contains salient information on explaining contemporaneous price impact and forecasting subsequent future returns.}

Our motivation arises from the fact that market participants can make trading decisions by observing the trade flows in the market. Previous works \citep{kyle1985continuous, kyle2011model} model the price formation at high frequency, and suggests that informed traders split large orders into many smaller orders in order to conceal their true purpose, while other market participants monitor order flows in the market in order to reach trading decisions. The development of high-performance trading systems has led to an astounding growth of high-frequency trading (HFT) and diversity of strategies \citep{hagstromer2013diversity}. In this world, the reaction time plays an important role because opportunities can be transient if not acted upon within micro-seconds, and even nano-seconds. High-frequency trading strategies include anticipating trade flow \citep{hirschey2021high} and preying on other market participants \citep{van2019high}. The questions we are interested in exploring concern whether certain trades, interacting with other trades in various different ways, contain useful information, and how they contribute to stock price movements, helping us shed light on the price formation mechanism at both short-term and long-term horizons. \update{To be specific, interaction refers to the fact that arrivals of trades may affect each other. Trades can occur in response to some events. Placements of trades can be initiated by the arrival of other trades or by changes in order imbalance, especially for HFT strategies based on observing order flows.}

We start with proposing the concept of \textit{co-occurrence of trades}, defined in \Cref{cooc_defination}, which offers  a tool to identify and group trades based on their interactions with other trades. For each given trade, we consider it to co-occur and interact with another trade if both trades are taking place close in time to each other. To define and quantify ''closeness", we pre-define a neighbourhood size $\delta$. If the time difference between two trades is lower than $\delta$, they are close to each other and they co-occur. Notice that the threshold $\delta$ is an important parameter, determining the set of trades that co-occur. However, there is no strict rule to set its value. Intuitively, considering a scenario where an HFT preys on an institutional trader and trades in response to institutional marketable orders, we aim to capture these interactions and classify such trades into a category of, for example, actively interactive trades. With this in mind, an appropriate choice should be greater than the round-trip latency plus the time for the HFT to detect and make trading decisions, which is usually undisclosed. Therefore, we experiment with multiple values of $\delta$, and compare and contrast the corresponding results. Note that $\delta$ should not be too large either, since a large neighbourhood is likely to incorporate irrelevant trades from the market. \update{To select the neighbourhood size, we first introduce a null model of completely random order arrivals. Then we select the $\delta$ that maximize the difference between the empirical co-occurrence of trades and the co-occurrence under the null model. We find that $\delta$ = 1 ms is an appropriate choice and use it for the empirical analysis in this study. In addition, we also make a comparison across different choices of $\delta$ values in \Cref{appendix:neighbourhood_size}.} 

Using trade co-occurrence, we decompose daily trade flows by classifying all the trades of all stocks into subgroups. Given a trade, we determine to which group it belongs by asking the following two questions: Does it interact with other trades? If yes, does it interact with only trades of the same stock as itself, only with stocks different from itself, or with both kinds? Depending on the answer, a trade will be placed into one or two classes, for which detailed rules are explained in \Cref{trade_condition}. After labeling all trades, we study the relations between returns and subgroups of trades.

We use order imbalance as a bridge connecting trade flows and stock returns, which has been thoroughly studied in the finance literature. An inventory paradigm \citep{stoll1978supply, spiegel1995intraday, chordia2002order} suggests that, in intermediated markets, a difference, or so-called \textit{imbalance}, between buyer-initiated and seller-initiated trades puts pressure on a market maker's inventory. In response, the market makers adjust inventories to maintain their market exposures, which drives the price to one direction. 

Next, at a daily level, we investigate the properties of aggregated order imbalance of each category of trades and their relation with individual stock returns during normal trading hours. Data exploration indicates that all categories of conditional, as well as the unconditional, order imbalance are positively auto-correlated. The conditional order imbalances (COIs) all have strong positive correlations with the original order imbalance. However, they are not necessarily highly correlated with each other. 

Our empirical results concentrate on the imbalance-return relations. By means of regression analysis, we discover positive and significant correlations between order imbalances and price changes within the same day. Furthermore, in comparison to a standard regression analysis, decomposing order flows leads to significantly higher adjusted $R^2$ in our multiple regression settings, which can be interpreted as better explanatory power in contemporaneous intraday open-to-close stock returns. To exploit predictability, we use the same regression analysis to fit order imbalances against future one-day ahead returns. In contrast to contemporaneous results, statistically significant relations only appear in order imbalance of isolated trades. Despite the absence of significant regression coefficients, we observe that order imbalances of non-isolated trades arrive closely with trades for other stocks, appear to have negative relations with future returns. On the contrary, imbalances of trades arrive together with only trades of the same stocks show weakly positive correlations. 

These associations are amplified in our subsequent portfolio analysis, as follows. 
We leverage these imbalances to build trading strategies.
In order to assess the economic value of the trade flow decomposition method, we construct signal-sorted portfolios using COIs as signals. In particular, if we make long/short decisions in alignment with the observed patterns in the predictive regressions, we attain profits in all of our portfolios, with the highest annualized Sharpe ratio reaching \update{1.79}. As a benchmark, we build portfolio investing in order imbalances without decomposition, for which the Sharpe ratio is negative. 

The remainder of this paper is organized as follows. \Cref{literature_review} outlines our contributions to the finance literature. In \Cref{definations}, we introduce the definitions of trade co-occurrence, trade flow decomposition and COIs. We start our empirical studies with describing data sources and conducting exploratory analysis in \Cref{data}. Subsequently, we uncover the relations between COIs and contemporaneous returns in \Cref{contamporaneous_analysis} and investigate the predictive power of COIs in \Cref{predictability}, and economic value of COIs in \Cref{economic_value}. \Cref{robustness} provides robustness analysis and additional empirical findings. Finally, in \Cref{conclusion}, we summarize the results and discuss our limitations and future research directions.

\section{Related literature} \label{literature_review}
This paper contributes to four strands of literature. First, our study exploits a new financial application of co-occurrence analysis, which is a statistical method proven to be powerful in spatial pattern analysis and widely used in the fields of biology \citep{gotelli2000null, mackenzie2004investigating, araujo2011using}, natural language processing (NLP) \citep{dagan1999similarity, kolesnikova2016survey}, computer vision \citep{galleguillos2008object, aaron2018image}, and others \citep{appel1998co, ye2017co}. So far, the applications of co-occurrence analysis in finance literature concentrate on studying stocks co-occurring in news articles. \citet{ma2011mining} construct networks from company co-occurrence in online news and use machine learning models to identify competitor relationships between companies. Recent studies, including \citet{guo2017news,tang2019news,wu2019deep}, build networks using stocks co-occurrence in news and employ them for tasks such as  return predictions and portfolio allocation. We contribute by originating the idea of trade co-occurrence. By directly applying the co-occurrence of stock trades, we establish that this technique is  beneficial for exploring and gaining insights from the financial market microstructure. 

Second, our research adds to the studies of interactions among trading activities in the market. In \citet{kyle1985continuous}'s model, market makers observe the aggregated order flows of informed and liquidity traders in the market to adjust their trading strategies. More aggressively, HFT traders can detect informed traders, such as institutions \citep{van2019high} and predict trade flows of others \citep{hirschey2021high}. Various theoretical models \citep{grossman1988liquidity, brunnermeier2005predatory, yang2020back}) are proposed for the interplay between high-frequency and institutional traders. \citet{van2019high} conduct an  empirical study on the Swedish stock market and discover that HFT participants intend to trade against wind when the institutional traders begin splitting large orders, and eventually trading in the same direction as the institutions. 

We contribute to this topic by proposing the idea of trade co-occurrence and provide empirical evidence that the co-occurrence of stock trades is not coincident. Rather than studying interaction among traders, we innovate trade co-occurrence as a tool to analyze interactivity at the individual  trade level. Our study of COIs conditional on co-occurrence shows that the interactions of trades at a granular level convey useful information on price formation.

Third, this paper contribute to the literature of order imbalance and price formation. According to pioneering researches, persistence in order imbalance can arise in two ways. Firstly, as the model by \citet{kyle1985continuous} states, traders intend to split large orders over time to minimize their market impacts, which leads to auto-correlated imbalances. Another source for order imbalance, as \citet{scharfstein1990herd} state, is the herd effect. To explore how order imbalance affects price changes, \citet{chordia2004order} propose a theoretical model to explain the positive relation between order imbalance and contemporaneous stock returns, arising from the market makers dynamically accommodating order imbalance. In addition, discretionary traders optimally splitting orders across days enables order imbalance to have strong positive auto-correlation and predictive power on future returns. Their empirical study, using daily data of stocks listed on New York Stock Exchange (NYSE) for a 10-year period from 1988 to 1998, confirms their theoretical results and shows that order imbalances have significant forecasting power on future returns. However, there is controversy on the predictability. For example, \citet{shenoy2007order} and \citet{lee2004order} find no significant predictive power of order imbalances. 

Although \citet{chordia2004order} do not differentiate trade flows, subsequent studies have shown that marketable orders, placed at different time, by different agents, with distinct properties can have different impacts on price changes. Most evidence stems from the Chinese market \citep{lee2004order, bailey2009stock, zhang2019order}, where private data of identification of trader types are available, and they find indications that order imbalances of institutional trade flows have higher pressure on prices than imbalances of individual traders. Same results are found in the US market by \citet{cox2021iso}'s recent study of S$\&$P 500 stocks during 2015 to 2016, which split trades into binary classes depending on whether or not they are inter-market sweeping orders, which are mainly adopted by institutions \citep{chakravarty2012clean}.

Our research complements these works by supplementing the study of order imbalances in the US market using data of the most recent period and proposing a novel method to decompose the unconditional trade flows without requiring an additional private data set. We show that order imbalances, without differentiating trades, no longer have forecasting power on future returns, which is evidence for an evolution of the market microstructure over the past decades \citep{chordia2002order,chordia2004order}. However, trade flows decomposed with our proposed method carry different information content, and their COIs do possess forecasting power. 

\update{Finally, this paper adds to the literature of trading strategies based on order flow signals  \citep{aldridge2013high}. Traders can boost the profitability of their strategies by analyzing the flow of orders in the market to improve their forecast signals, and gaining insight from the strategies of their competitors \citep{foster1996strategic, hirschey2021high}. Many previous studies have discovered that information derived from order flows, at a granular level, exhibits conspicuous predictive power on stock returns \citep{zhang2019deeplob, cont2021price, ait2022and, lucchese2022short}, and can thus be leveraged for developing profitable trading strategies  \citep{guilbaud2013optimal, bechler2015optimal, kolm2021deep, wang2021high}.  Along the same lines, order imbalances derived from order flow have been widely used in developing trading strategies \citep{cartea2015algorithmic}. \citet{chordia2004order} demonstrates the  profitability of order imbalances as trading signals. \citet{chang2012order} uses order imbalances to enhance the performance of daily price momentum strategies and generates significant returns.

We contribute to this field by proposing a method to analyze trade flows based on the time proximity of trade arrivals, and extract profitable trading signals form the aggregated trade flow. We leverage the derived COI-based signals to develop successful trading strategies, and showcase their profitability with rigorous backtest and robustness checks.}

\section{Co-occurrence of trades and trade flows decomposition} \label{definations}
\subsection{Co-occurrence of trades} \label{cooc_defination}
\begin{figure}[ht]
    \centering
    \includegraphics[width = \textwidth]{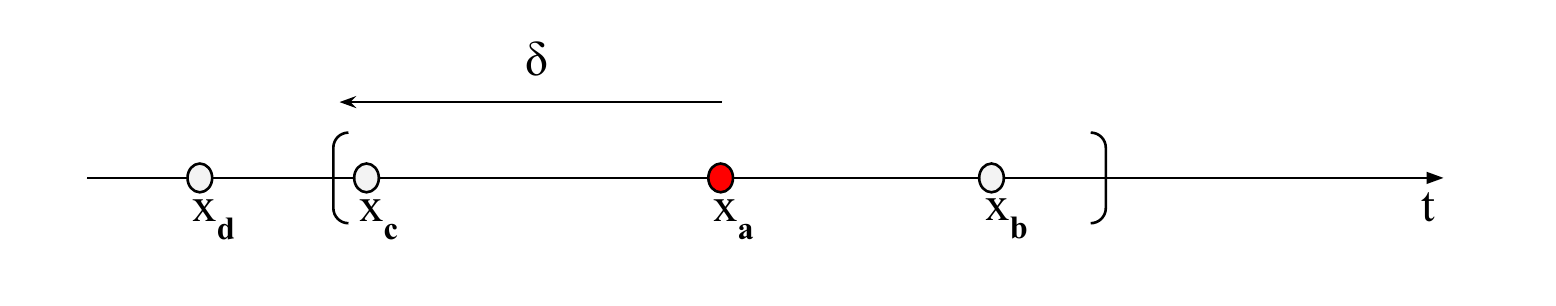}
    \vspace{-4mm}  
    \caption{Illustration of trade co-occurrence. This figure visualizes the idea of trade co-occurrence; given a user-defined neighbourhood size $\delta$, trade $x_b$ arrives within the $\delta$-neighbourhood of trade $x_a$, and thus they co-occur. In contrast, trade $x_d$ locates outside $x_a$'s neighbourhood, and thus the two trades do not co-occur. Both trades $x_b$ and $x_c$ co-occur with trade $x_a$, but they do not co-occur with each other} 
    \label{fig:trade_cooccurrence}
\end{figure}

We first introduce the definition of trade co-occurrence. For each trade $x_{a}$ occurring at time $t_a$, with a pre-specified $\delta$, every trade, other than $x_{a}$ itself, that arrives within time period $(t_{a} - \delta, t_{a} + \delta)$ is defined as having co-occurred with trade $x_{a}$. We define the threshold $\delta$ as the \textit{neighbourhood size}, and the set of all trades co-occurred with $x_{a}$ as $\delta$-neighbourhood of trade $x_{a}$, denote as $\mathbf{B}_{\delta}(x_a)$. \Cref{fig:trade_cooccurrence} sketches an example, where trade $x_a$ co-occurs with trades $x_{b}$ and $x_c$, while it does not co-occur with trade $x_{d}$. We note that co-occurrence is not an equivalence relation. It is perfectly possible for $x_a$ and $x_b$ to co-occur, and for $x_a$ and $x_c$ to co-occur, without $x_b$ and $x_c$ co-occurring. 

\subsection{Trade flow decomposition} \label{trade_condition}

\updateSecondRound{Based on co-occurrence, we next split the trades of every given stock into different classes characterized by their $\delta$-neighbourhood. We name this procedure as  \textit{trade flow decomposition}.}

\subsubsection{Definition} \label{sec:def_decomposition}
\updateSecondRound{For definition, we denote the set of trades of a given stock $i$ as $\mathbf{X}_{i}$. For a given universe of stocks, denoted as $\mathcal{S}$, our goal is to assign labels to each trade $x_{a} \in \mathbf{X}_{i}$ for every stock $i \in \mathcal{S}$.

In order to classify trades based on their time proximity with other trades in the market, we need to determine trades of which stocks other than stock $i$ shall be incorporated. Thus, we introduce a fixed set of stocks as a customized market index, denoted by $\mathcal{M}$, whose trades are also considered when labeling trades of stock $i$. Then we define the set of trades, $\mathbf{M}$, as a representative of the market, \mc{referred to} as the \textit{market set}, that is $\mathbf{M} = \cup_{j \in  \mathcal{M}} \mathbf{X}_{j}$, for all stocks $j \in \mathcal{M}$.

Note that the stock $i \in \mathcal{S}$, whose trades we aim to label, may or may not be in the \textit{market set}, $\mathcal{M}$. Therefore, for each stock $i \in \mathcal{S}$, we construct a reference set, $\mathbf{M}_{-i} = \mathbf{M} - \mathbf{X}_{i}$, which contains all trades of stocks other than stock $i$, in the market set. Finally, every trade $x_a \in \mathbf{X}_{i}$ is equipped with the set $\mathbf{B}_{\delta}(x_a)$ of trades in its neighbourhood. }


%

%

%
%

%

\begin{figure}[t]
    \centering
    \includegraphics[width = \textwidth]{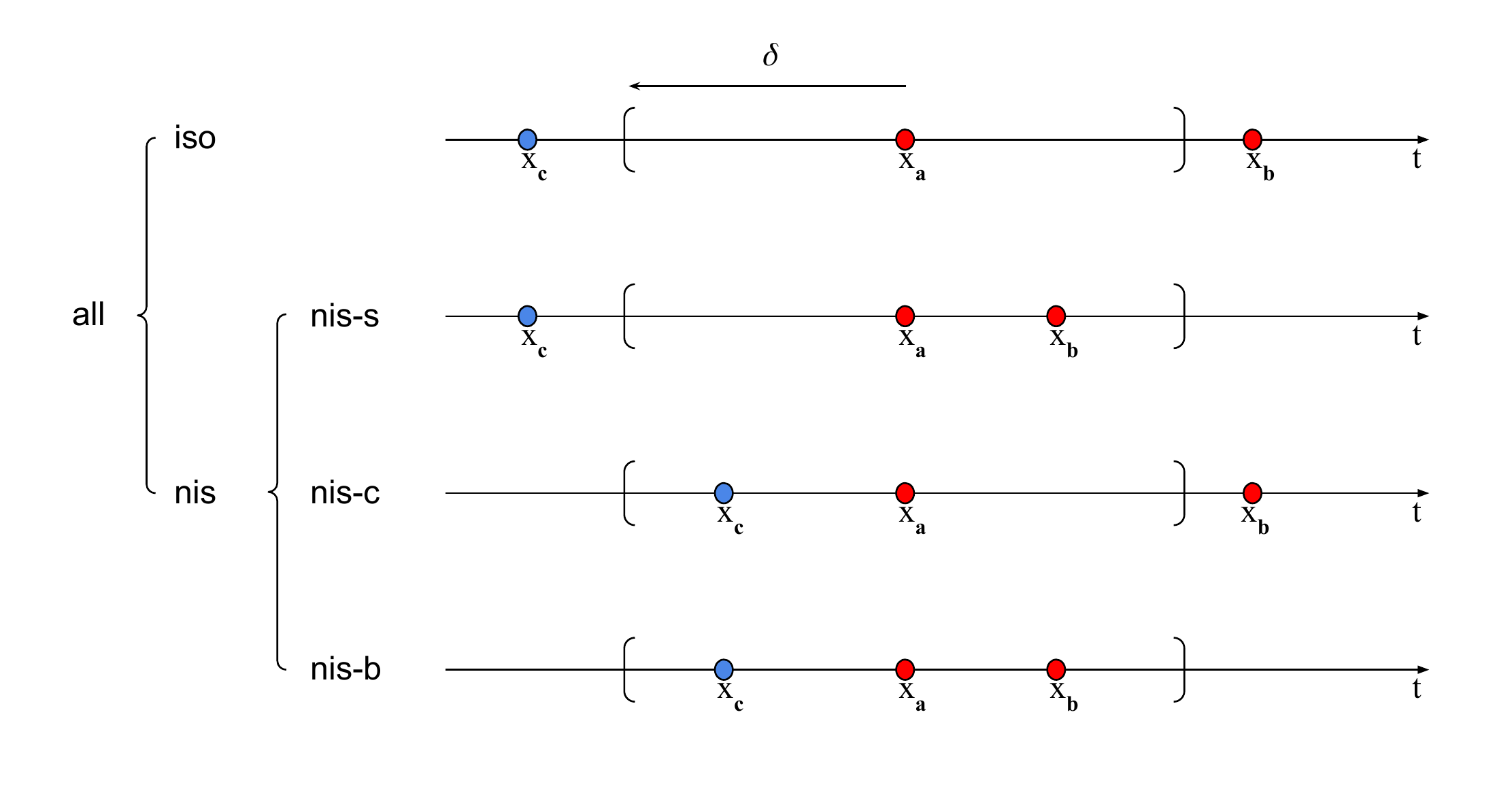}
    \caption{Illustration of trade types, conditioning on co-occurrence. We showcase the distinct categorical labels of trade $x_{a}$. Color indicates the stock corresponding to a trade. Thus, $x_b$ is for the same stock as $x_a$, while $x_c$ is for a different stock. First line: 
    $x_{a}$ is an isolated (iso) trade with empty $\delta$-neighbourhood; second to fourth lines: $x_{a}$ is a non-isolated (nis) trade with nonempty $\delta$-neighbourhood; second line: $x_{a}$ is a non-self-isolated (`nis-s') trade with only other trades for the same stock in its $\delta$-neighbourhood;  third line: $x_{a}$ is an non-cross-isolated (`nis-c') trade with only other trades for the different stocks in its $\delta$-neighbourhood; last line: $x_{a}$ is a non-both-isolated (`nis-b') trade with both other trades for the same and different stocks in its $\delta$-neighbourhood.}
    \label{fig:conditional_cooccurrence}
\end{figure}

\updateSecondRound{With these sets, we formally define the trade flow decomposition by assigning each trade of stock $i$ \mc{to one or two of} five categories,} with the protocol illustrated in \Cref{fig:conditional_cooccurrence}. Initially, we partition all trades into two groups, isolated (iso) and non-isolated (nis) trades, defined as follows
\begin{enumerate}
\item  \textit{isolated} (iso): \updateSecondRound{A trade, $x_a \in \mathbf{X}_{i}$, is labelled as \textit{isolated} if it does} not co-occur with any other trade, \updateSecondRound{that is $\mathbf{B}_{\delta}(x_a) \cap \mathbf{X}_{i} \cap \mathbf{M}_{-i} = \varnothing$};
\item \textit{isolated} (nis): \updateSecondRound{The trade is labelled as \textit{non-isolated} if there are other trades of the same stock, or trades of other stocks in the market index, $\mathcal{M}$, in its neighbourhood, that is $| \mathbf{B}_{\delta}(x_a) \cap \mathbf{X}_{i} \cap \mathbf{M}_{-i} | \geq 1$, where $| \boldsymbol{\cdot} |$ denote the cardinality of a set}.
\end{enumerate}

We further decompose the non-isolated trades according to properties of the trades within their $\delta$-neighbourhood. Each non-isolated trade $x_a \in \mathbf{X}_{i}$ can be classified into one of the following three categories
\begin{enumerate}\addtocounter{enumi}{2}
\item  \textit{non-self-isolated} (nis-s): the $\delta$-neighbourhood of trade $x_a$ contains \textbf{only} trades (at least one) of the \underline{same} stock as the one from trade $x_a $, \updateSecondRound{ that is $| \mathbf{B}_{\delta}(x_a) \cap \mathbf{X}_{i} | \geq 1$ and $| \mathbf{B}_{\delta}(x_a) \cap \mathbf{M}_{-i} | = 0$}; 

\item  \textit{non-cross-isolated} (nis-c): the $\delta$-neighbourhood of trade $x_a$ contains \textbf{only} trades of stocks which are \underline{different} than the stock corresponding to trade $x_a$, \updateSecondRound{ that is $| \mathbf{B}_{\delta}(x_a) \cap \mathbf{X}_{i} | = 0$ and $| \mathbf{B}_{\delta}(x_a) \cap \mathbf{M}_{-i} | \geq 1$};

\item   \textit{non-both-isolated} (nis-b): the $\delta$-neighbourhood of trade $x_i$ contains \textbf{both} at least one trade of the \underline{same} stock, and at least one other trade of a \underline{different} stock, \updateSecondRound{ that is $| \mathbf{B}_{\delta}(x_a) \cap \mathbf{X}_{i} | \geq 1$ and $| \mathbf{B}_{\delta}(x_a) \cap \mathbf{M}_{-i} | \geq 1$}. 
\end{enumerate}

These three classes form a partition of the set of non-isolated trades, as illustrated in \Cref{fig:conditional_cooccurrence}. We refer to this process of separating trades into categories as \textbf{trade flow decomposition}.

\update{
\subsubsection{Motivation and generating mechanism} \label{sec:mechanism}
The motivation behind our decomposition is to separate the trade flows of different types of market participants. Each trade flow should be dominated by certain types of traders, and is thus expected to have distinct impact on  stock returns. 
\begin{enumerate}
    \item iso: Informed traders, for example financial institutions with access to sophisticated private alphas and infrastructure, 
    tend to hide their trading purposes. When they are successful, their trades should neither follow nor be followed by other trades, thus becoming locally isolated. We expect this type of trade flow to exhibit significant price impact and to be consistent with long-term price changes.
    
    \item nis: Excluding informed trade flow, we expect this type of flow to have negative relationship with future price changes. However, the majority of the market participants should not have insider information, rendering this type of trade flow to have larger trading volume. Therefore, it should have considerable impact on contemporaneous stock returns. 
    
    \item nis-s: HFT traders, who anticipate or identify the trades placed by the aforementioned informed traders, can front-run or prey on those trades. Therefore, these types of trades along with unsuccessfully hidden  trades from informed trades are likely to co-occur. We thus expect this type of order flow to have the same direction of price pressure as the iso flow, but with less impact and consistency.
    
    \item nis-c: Traders who run market neutral strategies or trade baskets of stocks will rebalance when their positions in other stocks change, or trade multiple stocks simultaneously. This type of trade flow should capture most of this rebalancing (for example, updating positions of index constituents in index arbitrage strategies). We expect this mass rebalancing behaviour to exhibit both permanent and transient impact, and to lead to price mean reversion in the next period.
        
    \item nis-b: When the market intensity suddenly rises, for example, due to release of news concerning macroeconomic events or increased trading activity around market opening and closing sessions, trading volumes increase across all stocks, leading to the arrival of such types of trades. Potential overreaction to such news events could result in mean reversion of future prices. 
\end{enumerate}

In addition, we assume there exists noise traders who are likely to be classified in any of these trade flow categories, and who choose their trading direction randomly. To assess the overall price impact, we calculate the order imbalance of each type to obtain its net price pressure; see details described in the following subsection. To closely examine the mechanism, we perform an analysis on the relationship between order imbalances of decomposed trade flows (COIs), across both contemporaneous and future stock returns. Therefore, our hypotheses are two-fold
\begin{enumerate}
    \item all types of COIs have significant positive relation with contemporaneous returns;
    \item order imbalances of iso and nis-s trade flows are positively related with future returns, while COI of the other types are negatively correlated with future returns.
\end{enumerate}
It is important to clarify that without client order ID data and information on the type of strategy from which the individual orders originate, it is challenging to identify and establish the generating mechanisms behind each type of decomposed order flow.}

\subsection{Conditional order imbalance}
With the decomposition of trade flows, we proceed to study the price impact of trades with different characteristics. A bridge connecting trading activities and price changes is given by the order imbalance quantity, defined as the normalized difference between the volume of buyer- and seller-initiated trades \citep{chordia2004order}. For a given stock $i$, we derive conditional daily order imbalances, as follows
\begin{equation}
   COI_{i, t}^{type} = \frac{N_{i, t}^{type, buy} - N_{i, t}^{type, sell}}{N_{i, t}^{type, buy} + N_{i, t}^{type, sell}},
\end{equation} 
where $N_i^{buy, type}$ and $N_i^{sell, type}$ denote the total number of market buy orders and market sell orders of stock $i$ in day $t$ respectively. If the denominator is 0, which happens when there are no trades of a certain type, we define the COI in this case to be 0. We consider six
types of COIs and the superscript $type$, which takes a value in $\{$all, iso, nis, nis-s, nis-c, nis-b$\}$, indicates the group of trades used to calculate the imbalance. Note that the `all' label  
corresponds to using the entire universe of trades without decomposing based on trade co-occurrence. Thus,  the 
`all' COI is the same as order imbalance in the number of transactions, scaled by total transactions, studied by \citet{chordia2004order}.

\section{\update{Empirical selection of $\delta$, existence of co-occurrence,  and exploratory data analysis}} 
\label{data}

\update{In this section, we propose an empirical approach for choosing the parameter $\delta$, and we showcase the existence of co-occurring trades in the market. We start with a brief description of the data employed in our study. We then provide empirical evidence that setting $\delta = 1$ ms is an appropriate choice. For further details of different values of $\delta$, we 
refer the reader to \Cref{appendix:neighbourhood_size}.} 
Moreover, we uncover salient patterns of trade co-occurrence through exploratory analysis. Furthermore, we show that the resulting order imbalances of the decomposed trade flows are only weakly correlated with each other, which indicates that the trade decomposition we propose is meaningful.

\subsection{Data source and preprocessing}
Our study is based on 457 US stocks during the period from 2017-01-03 to \update{2020-12-31}. The selected stocks are those companies included in Standard $\&$ Poor's ($S\&P$) 500 index for which both order book data and price data is available over the entire sample period. \update{\Cref{tab:sum_stats_sample_stocks} provides of brief summary of the stocks.}

\subsubsection{Limit order book data}
We obtain limit order book data from the LOBSTER database \citep{huang2011lobster}, which provides detailed records of limit orders for all stocks traded in the NASDAQ exchange. The records include limit order submissions, cancellations and executed trades, indexed by time with precision up to nanoseconds. For each stock on each trading day, a record contains the time stamp, event type (submissions/cancellations/executions), direction (buy/sell), size and price for a limit order event. By filtering for  limit order executions and reversing their directions, we infer the buyer- and seller-initiated trades, e.g. execution of a limit buy order implies placement of a market sell order/marketable limit sell order. Noticing that a large market order simultaneously consumes multiple existing limit orders, we merge inferred trades with identical timestamps. Given LOBSTER's high time resolution, we assume different trades cannot have exactly the same timestamps.

\subsubsection{Prices and returns}
We acquire daily price data for our stock universe under consideration, from the Center for Research in Security Prices (CRSP) 
database, and calculate daily open-to-close logarithmic returns as 
\begin{equation}
    R_{i, t} = \log \frac{P_{i, t}^{Close}}{P_{i, t}^{Open}},
\end{equation}
where $P_{i, t}^{Open}$ and $P_{i, t}^{Close}$ are daily open and close prices of stock $i$ on day $t$. To alleviate the effect of the market  component, we also consider \textit{market excess returns} in this study,  denoted as $r_{i,t}$, calculated as follows
\begin{equation}
    r_{i, t} = R_{i, t} - R_{SPY, t},
\end{equation}
where $R_{SPY, t}$ is the daily return of SPY ETF, which tracks the S$\&$P 500 index. For simplicity, here we assume all stocks have the same market \textit{beta equal to 1.} 

\update{In addition, we collect factor data from Kenneth R. French's online
Data Library.\footnote{We obtain the data of factors from Kenneth French's website. \\
\href{https://mba.tuck.dartmouth.edu/pages/faculty/ken.french/Data_Library/f-f_factors.html}{https://mba.tuck.dartmouth.edu/pages/faculty/ken.french/Data$\_$Library/f-f$\_$factors.html}}
These include daily returns of the market factor (MKT), size factor (SMB), value factor (HML), profitability factor (RMW), investment factor (CMA) \citep{fama1992cross, fama1993common, fama2015five}, and momentum factor \updateSecondRound{(MOM)} \citep{jegadeesh1993returns, carhart1997persistence}. 
}

\updateSecondRound{
\subsection{Universe of stocks and the representative of the market}
In the empirical research, we classify trades of stocks in a universe comprising of 457 constituents of the S\&P 500 index. For simplicity, we also use the set of all trades of the same 457 stocks as representative of the market. According to the definition in \Cref{sec:def_decomposition}, that is $\mathcal{S} = \mathcal{M}$. Therefore, the reference set of each stock $i$, $\mathbf{M}_{-i} = \mathbf{M} - \mathbf{X}_{i}$, consists of trades of 456 stocks other than itself. We are aware of that the labels of trades can depend on the market set, and discuss the selection of market indices in \Cref{subsec:universe_of_stocks}.

}

\subsection{\update{Null model: co-occurrence probabilities under complete randomness}}
With order book data, we first answer the following fundamental questions. Do trades really co-occur or are their arrivals simply random and independent of each other? Does our trade flows decomposition capture a signal? In this section, we develop a null model under the assumption of completely random order arrival. 

We assume that, for stock $i$, the arrivals of trades within a time interval of length $T$, follow independent Poisson processes with the same intensity $\lambda_{T}$. Let $N_i$ denote the number of trades of stock $i$ in $[0,T]$.
Conditional on  $N_{i} =n_i$,  the arrival time of the $n_i$ trades are independent and follow a uniform distribution on $[0,T]$.  Hence, for each trade, the probability that another trade falls in its $\delta$-neighbourhood during the time period $T$ is 
\begin{equation}
    p = \frac{2 \delta}{T}.
\end{equation}

Next, we derive the probabilities of different types of trade flows, as follows
\begin{equation}
    \begin{split}
         &\mathbb{P}_{i}^{\delta} (\textit{iso}) = (1 - p) ^ {(N_{i} + N_{-i} - 1)}, \\
         &\mathbb{P}_{i}^{\delta} (\textit{nis}) = 1 - (1 - p) ^ {(N_{i} + N_{-i} - 1)}, \\
         &\mathbb{P}_{i}^{\delta} (\textit{nis-s})= [1 - (1 - p) ^ {N_{i} - 1}] (1 - p) ^ {N_{-i}}, \\
         &\mathbb{P}_{i}^{\delta} (\textit{nis-c})= (1 - p) ^{N_{i} - 1} [1 - (1 - p) ^ {N_{-i}}], \\
         &\mathbb{P}_{i}^{\delta} (\textit{nis-b})= [1 - (1 - p) ^ {N_{i} - 1}] [1 - (1 - p) ^ {N_{-i}}],
    \end{split}
\end{equation}
where $N_{-i}$ denotes the number of trades for all stocks in the market other than stock $i$. \update{In particular, for each stock $i$ in our sample universe of 457 stocks, $N_{-i}$ is the total number of trades of the remaining 456 stocks. }

\update{
\update{\subsection{Choice of neighbourhood size  $\delta$}}
The definition of trade co-occurrence and classification of individual trades depends on the choice of the neighbourhood size $\delta$. When considering the extreme case of $\delta = 0$, all trades are isolated. As we progressively increase $\delta$, an isolated trade turns into one sub-type of non-isolated trades. Meanwhile, both non-self-isolated and non-cross-isolated trades can only become non-both-isolated. Eventually, when $\delta$ is large enough, all trades are non-isolated; to be specific, they all become non-both-isolated. Hence, with the value $\delta$ increasing, the number of isolated trades decreases and the numbers of non-isolated and non-both-isolated trades increase monotonically. Thus, the quantities of non-self-isolated and non-cross-isolated trades initially increase; after reaching their respective maximum, they begin to decrease. We are aware that the choice of $\delta$ may depend on the specific task at hand, and the optimal value can vary; however, we propose a simple approach to select  $\delta$ for the following empirical study in this paper. The intuition is straightforward; we choose a $\delta$ which maximizes
the average distance, weighted by the empirical percentage of each type of trades, between null probabilities and empirical proportions. For simplicity, 
the same value of $\delta$ is shared across all stocks. We 
report the resulting average  distance in \Cref{tab:weighted_difference_delta}.

The first step is to derive the probabilities for each stock under complete randomness. As the intraday intensities are not constant, we thus calculate the probabilities for every \update{5 minutes ($T = 5$ min), which leads to 78 intervals}, and consider their averages (weighted by the intensities), as the final daily probabilities. We then also compute the empirical probabilities. We search over 8 values of neighbourhood size, $\delta \in \{ 0.05 $ ms$, 0.075 $ ms$, 0.125 $ ms$, 0.25 $ ms$, 0.5 $ ms$, 1 $ ms$, 5 $ ms$, 50 $ ms$ \}$, and plot their intraday null and empirical probabilities in \Cref{fig:cooc_prob_5min}. \Cref{tab:weighted_difference_delta} shows the average distance for the candidate $\delta$s. The maximum distance of 0.14 is achieved at $\delta$ = 1 ms.
\begin{figure}[h]
    \centering
    \includegraphics[width = \textwidth]{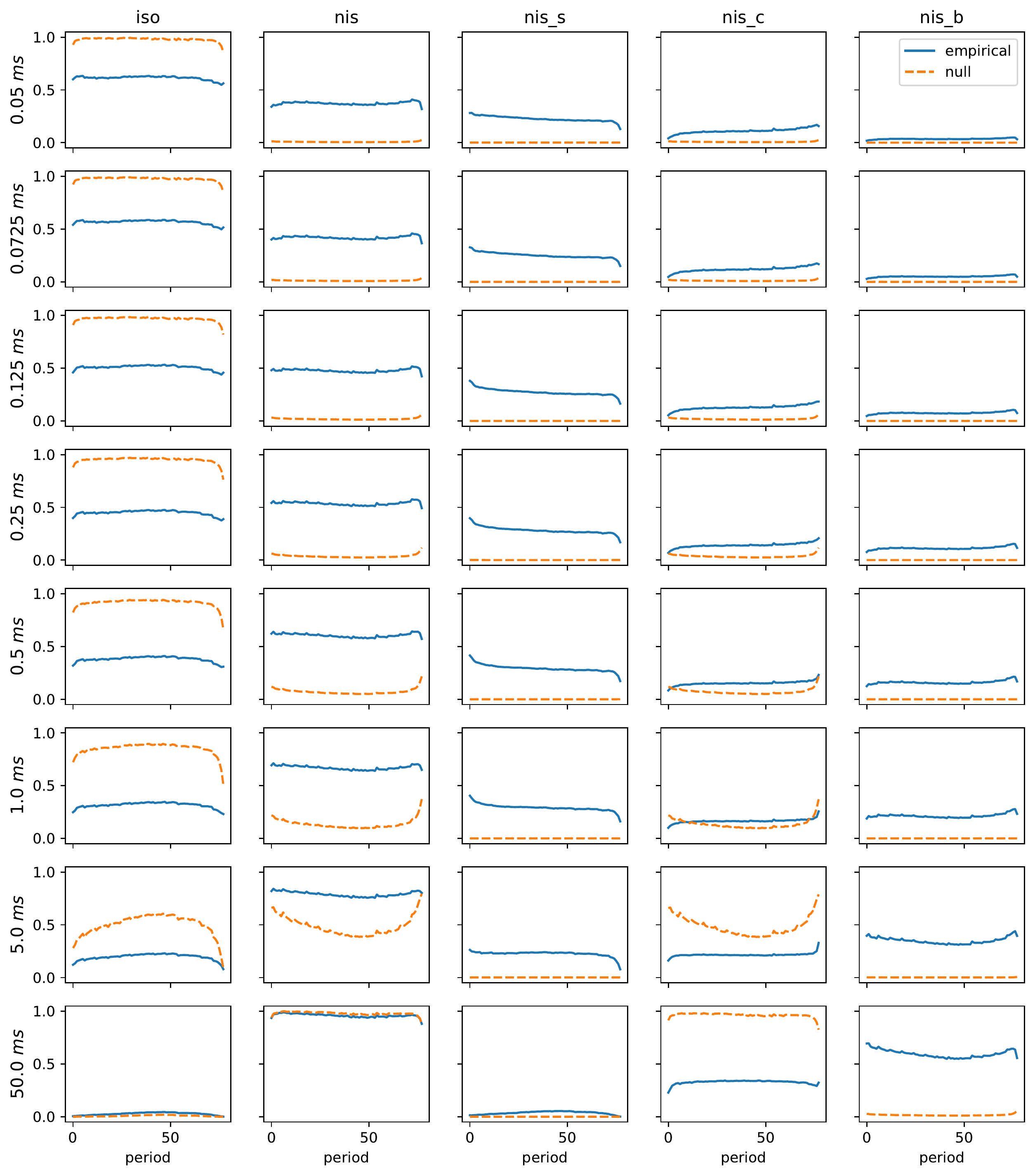}
    \caption{Co-occurrence probability: null vs. empirical. \\ The table plots the null and empirical probabilities of each type of trades, over 5 minute intervals, averaged over stocks and days, for selected values of $\delta$s.}
    \label{fig:cooc_prob_5min}
\end{figure}

\begin{table}[htp]
    \caption{Difference between null and empirical probability. \\ This table reports the average weighted distance for different values $\delta$ indicated in the columns.} 
    \label{tab:weighted_difference_delta}
        \begin{tabularx}{\textwidth}{l*{9}{Y}}
        \toprule
          $\delta$ (ms)    &  0.05  & 0.0725 &  0.125  &  0.25   &  0.5   & 1    &  5     & 50   \\ 
         \midrule
        average weighted distance &  0.09  & 0.10   &  0.11   &  0.12   &  0.13  & \bftab 0.14  & 0.11  &   0.12   \\
        \bottomrule
    \end{tabularx}
\end{table}

\subsection{Existence of co-occurrence}
By comparing the theoretical co-occurrence probabilities \citep{donges2016event} under the null model and the empirical values derived from data, we confirm the existence of co-occurrence among stock trades at the level of \update{1 millisecond}, supporting the idea that the overall trading volume has a strong cross-asset interaction component. From an economic perspective, this is perhaps to be expected, given the large presence in 
current markets of index-arbitrage traders who \textbf{simultaneously} trade an index ETF against a basket of constituents.
\begin{table}[t]
    \caption{Null and empirical probability of each type of trade flows. \\ This table shows the percentage, averaged over both days and stocks, of each type of trades under the null model and from the real data respectively, \update{with $\delta$ = 1 ms and T = 5 min}.}
        \begin{tabularx}{\textwidth}{l * 3 {Y}}
        \toprule
                 &   Null (\%)    &  Empirical (\%) \\
        \midrule
        iso      &   81.35        &  28.55          \\
        nis      &   18.65        &  71.45          \\
        nis-s    &   0.09         &  29.75          \\
        nis-c    &   18.53        &  17.27          \\
        nis-b    &   0.03         &  24.43          \\ 
        \bottomrule
    \end{tabularx}
    \label{tab:theoretical_prob}
\end{table}

\Cref{tab:theoretical_prob} shows the null and empirical daily probabilities averaged over time and stocks. Given the very small neighbourhood size \update{($\delta=1$ ms), $81.28\%$ of trades} should be isolated if there is no co-occurrence. However, there are only \update{$28.55\%$} isolated trades in the market. In conclusion, there is empirical evidence that the notion of trade co-occurrence captures a latent signal. This serves as motivation to further decompose trade flows and study them individually.}

\subsection{Summary statistics of trades}
\begin{figure}[t]
    \centering
    \includegraphics[width = \textwidth]{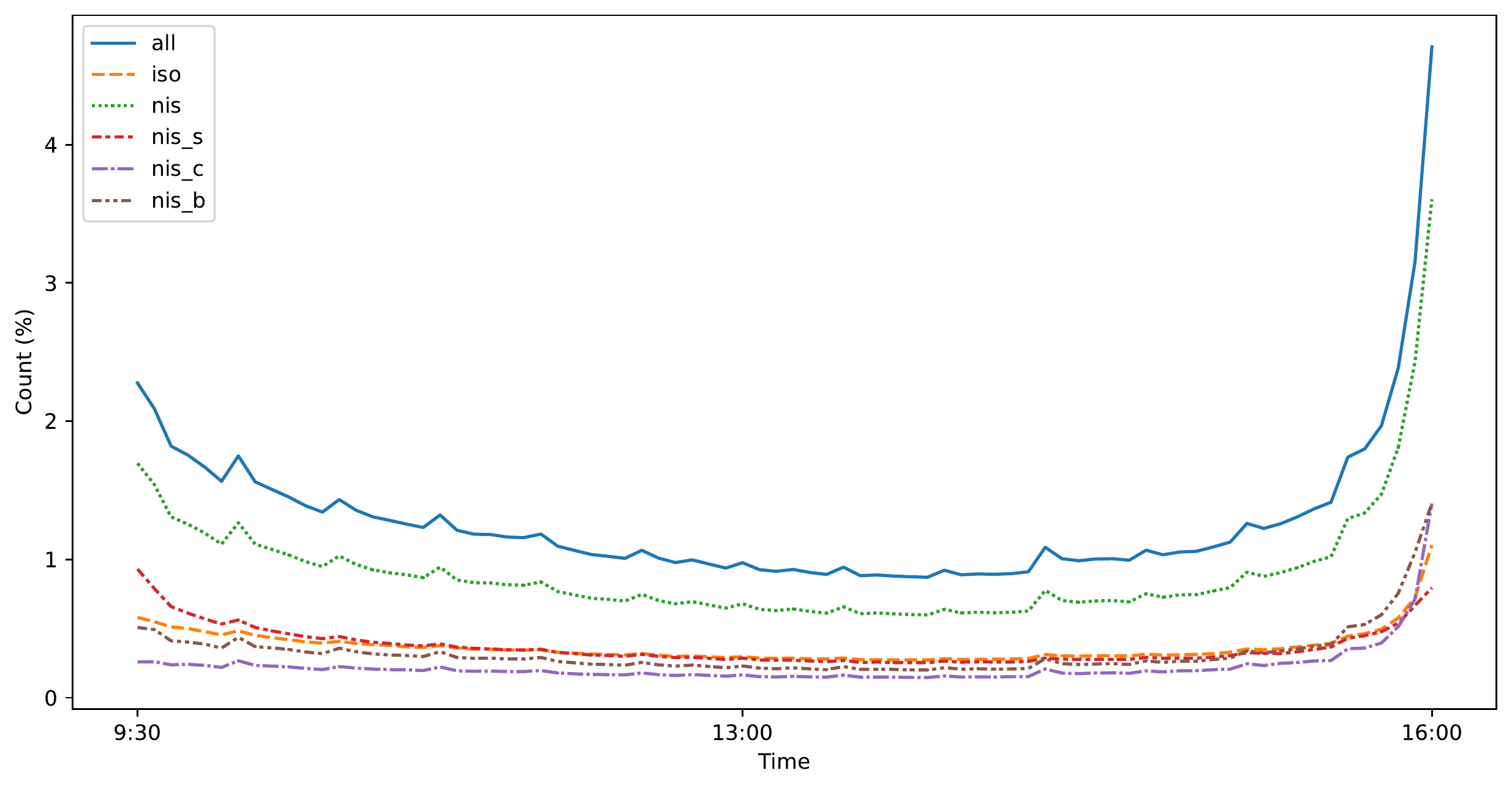}
    \caption{Intraday distributions of the number of each type of trades. \\ We calculate numbers of each types of trades in percentage of the total number of trades, for non-overlapping \update{5-minute} intervals
    during normal trading hours from 9:30 to 16:00 for all stock over the period from 2017-01-03 to \update{2020-12-31}. This figure plots intraday \update{5-minute} counts of different types of trades, averaged over both time series and cross-section.}  
    \label{fig:intraday_trade_dist}
\end{figure}

After building our data set of trades, we label every trade with its corresponding type. \Cref{fig:intraday_trade_dist} illustrates the intraday distributions of different types of trades. A summary of the data is presented in \Cref{tab:sum_stats_trades}; the chosen neighbourhood size for co-occurrence is \update{1 millisecond ($\delta = 1$ ms)}. The table shows descriptive statistics of the raw data, where each number is calculated by averaging daily time series and then considering the cross-sectional mean, median or standard deviation over all stocks. On average, isolated trades account for \update{$28.55\%$} of the total number of transactions, while the majority of trades are non-isolated in one of the three defined types (nis-s, nis-c, or nis-b). Approximately half of the non-isolated trades, \update{$29.75\%$} of all trades, are non-self-isolated. The mean proportions of non-cross-isolated and non-both-isolated trades are \update{$17.27\%$ and $24.43\%$}, respectively. The large standard deviation for the number of trades could be seen as an indication that the population is heterogeneous. The percentages of different groups of trades in terms of volumes, which are very similar to those reported in \Cref{tab:sum_stats_trades}. With this in mind, it is reasonable to concentrate on the count of trades as a liquidity measure.

\begin{table}[t]
    \caption{Summary statistics for all groups of trades. \\ This table documents the time-series average of the daily cross-sectional statistics of each type of trades. Our data include records of trades within normal trading hours of 457 stocks from 2017-01-03 to \update{2020-12-31}.} \label{tab:sum_stats_trades}
        \begin{tabularx}{\textwidth}{l*{4}{Y}}
        \toprule
        & Mean   & Median    & Std. dev. \\
        \cmidrule(lr){2-4} 
        
        Number of trades           &  3906.00 &  3010.85 &    3425.05 \\
        Percentage of iso trades   &    28.55 &    28.31 &       3.91 \\
        Percentage of nis trades   &    71.45 &    71.69 &       3.91 \\
        Percentage of nis-s trades &    29.75 &    29.34 &       6.34 \\
        Percentage of nis-c trades &    17.27 &    17.29 &       3.64 \\
        Percentage of nis-b trades &    24.43 &    23.95 &       4.69 \\
    \bottomrule
    \end{tabularx}
\end{table}

Highlighting the empirical fact that the trading activity is higher at the start and end of a trading day, \Cref{fig:intraday_trade_dist} plots the intraday distributions of trades, revealing slightly different temporal behaviours of different trade types. The plot exhibits the number of each type of trades over every half hour, with the $y$-axis indicating percentages of the total number of trades. We observe that all types of trades increase drastically in the last half an hour. It is noteworthy that, after the decomposition, the flow of isolated trades is smoother than the flow of non-isolated-trades, with a lower slope for the last-half-hour climb. By further separating the sub-types of non-isolated trades, we find that non-self-isolated trades contribute more at the start of a day, while the line of other two types are flat except at the end of days. 

\subsection{Descriptive statistics of order imbalances}
With trades labeled according to their co-occurrence types, we compute daily order imbalances and report descriptive statistics in \Cref{tab:sum_stats_coi}. Panel A documents summary statistics of each category of order imbalance, averaged over time and stocks. Overall, the average unconditional order imbalances are negative. After the decomposition, the isolated and non-self-isolated order imbalances tend to be negative, with both higher means and variances compared to their unconditional counterparts. In contrast, the means of non-cross-isolated and non-both-isolated imbalances are positive, but with even higher variance. Hence, our study essentially constructs features with different behaviours by conditioning on the co-occurrence of trades. 
However, the standard deviations are much larger than the means, so statistically, the means are not significantly different from zero. Hence the means can only be taken as a very weak indication of a potential signal.

Panel B presents average partial autocorrelations of each type of order imbalance. It can be seen that all the order imbalances are positively auto-correlated. The lag 1 auto-correlations for COIs are substantial. Among the conditional imbalances, the non-cross-isolated order imbalance, corresponding to trades that closely co-occur with trades of other stocks in the market, has relatively higher auto-correlation. In contrast, the  auto-correlation for the order imbalance from non-self-isolated trades is comparatively lower. These partial auto-correlations decay drastically with  increasing lags. 

\begin{table}[t]  
    \caption{Summary statistics for all groups of trades and order imbalances. \\ This table shows the summary statistics of COIs from 2017-01-03 to \update{2020-12-31} for the selected 457 stocks. Panel A documents the mean, median and standard deviations of COIs. Panel B presents the partial autocorrelations of COIs, averaged over all stocks.} \label{tab:sum_stats_coi}
        \begin{tabularx}{\textwidth}{l * 4 {Y}}
        \toprule
        \multicolumn{4}{l}{\textbf{Panel A: Statistics of daily order imbalances}} \\
        \midrule
                     & Mean      & Median     & Std. dev.   \\
         \midrule
        all    & -0.0012 & -0.0014 & 0.1144 \\
        iso    & -0.0095 & -0.0091 & 0.1561 \\
        nis    & 0.0030  & 0.0022  & 0.1196 \\
        nis-s  & -0.0058 & -0.0054 & 0.1445 \\
        nis-c  & 0.0161  & 0.0130  & 0.1622 \\
        nis-b  & 0.0119  & 0.0091  & 0.1724 \\

    \end{tabularx}

    \begin{tabularx}{\textwidth}{l * 4 {Y}}
        \toprule
        \multicolumn{4}{l}{\textbf{Panel B: Average partial autocorrelation of order imbalances}} \\
        \midrule

         lag         &  1      &  2       &  3     \\
         \midrule
        all         & 0.274 & 0.093 & 0.043 \\
        iso         & 0.243 & 0.097 & 0.048 \\
        nis         & 0.273 & 0.095 & 0.045 \\
        nis-s       & 0.226 & 0.092 & 0.049 \\
        nis-c       & 0.297 & 0.116 & 0.062 \\
        nis-b       & 0.254 & 0.099 & 0.056 \\
        \bottomrule
    \end{tabularx}
    
\end{table}

\begin{figure}[t]
    \centering
    \includegraphics[width = 0.5\textwidth]{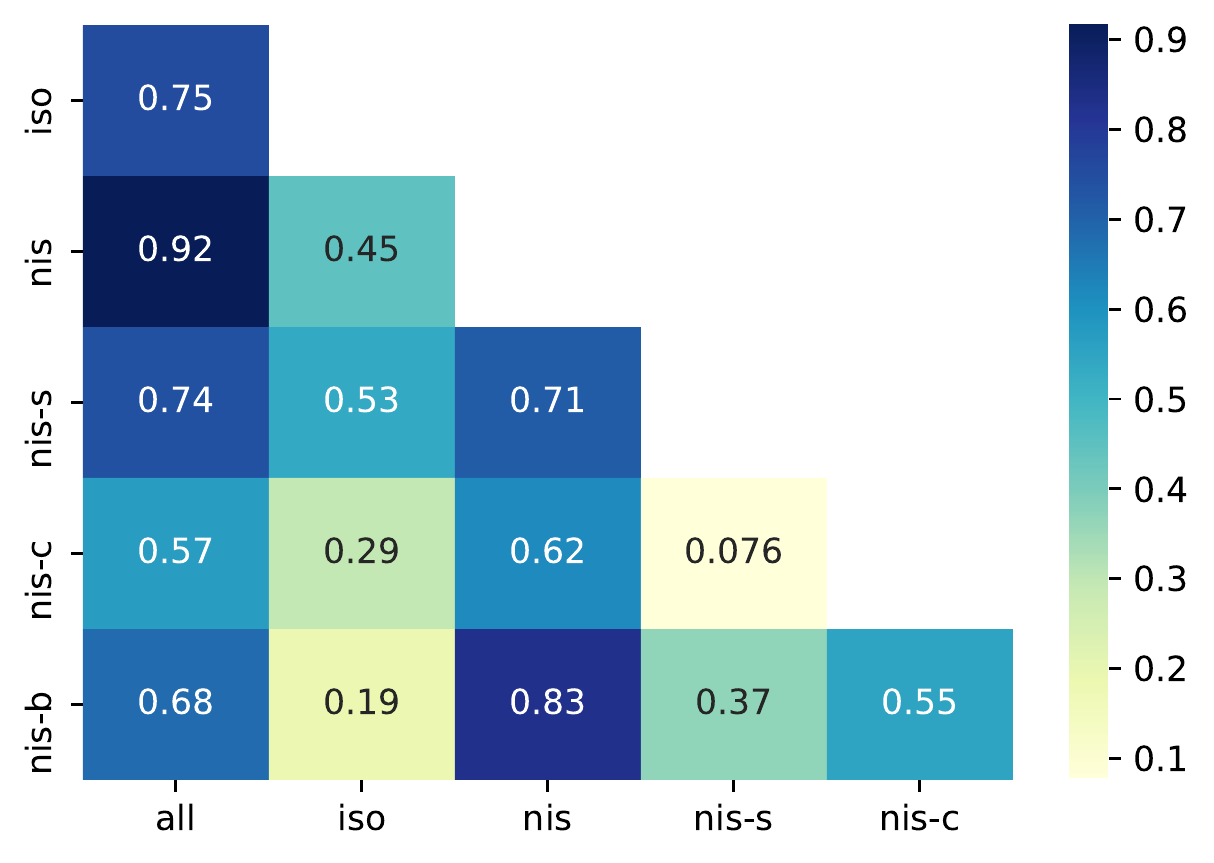}
    \caption{Pearson correlation of order imbalances. 
    For each type of order imbalance, we first consider the vector of daily values during 2017-01-03 to \update{2020-12-31}, then compute the correlation matrix, and finally average across all stocks.
    }
    \label{fig:oi_corr_plot}
\end{figure}

\Cref{fig:oi_corr_plot} shows the Pearson correlations, averaged over all stocks, of COIs, with \update{$\delta = 1$ millisecond}. All types of order imbalances are positively correlated with each other while the strengths are different and can be fairly low. An exception is the unconditional order imbalance, which is strongly associated with every other type. The correlations between isolated imbalance and non-isolated imbalance, as well as its sub-types, are low. 

As expected, conditioning on isolation and non-isolation produces distinct features. Furthermore, the three order imbalances obtained by decomposing non-isolated trades are also strongly correlated with the aggregated non-isolated order imbalance, but weakly correlated with each other. Upon exploring their relations in more detail, we find that the non-self-isolated order imbalances derived from orders which are not co-traded with other stocks in the market, are relatively more correlated with isolated order imbalances. In contrast, the order imbalances of non-cross-isolated and non-both-isolated trades, which are more connected with the market, are less correlated with the  isolated and non-self-isolated order imbalances. Therefore, we are confident that the decomposed order imbalances are distinguishable features,
with all pairwise correlations smaller than 0.6,  
that they can reveal insights about structural properties of the equity market which cannot otherwise be inferred by looking at the aggregated order flow.

\section{Contemporaneous price impact of conditional order imbalances} \label{contamporaneous_analysis}
To assess the contemporaneous effects of each type of order imbalance on contemporaneous returns, we employ the following \update{panel regression
\begin{equation} \label{contemporaneous_regression_equation}
\begin{split}
    r_{i, t} = \; &\alpha + \sum_{\rho \; \in  \; {Types} } \beta_{\rho} COI_{i, t} ^{\rho} + \beta_{\sigma} \sigma_{i, t} + \beta_{v} vol_{i, t} \\ 
    &+ \beta_{b} MKT_{t} + \beta_{s} SMB_{t} + \beta_{h} HML_{t} + \beta_{r} RMW_{t} + \beta_{c} CMA_{t} + \beta_{m} MON_{t} + \epsilon_{i, t},
\end{split}
\end{equation} 
where $r_{i,t}$ is the return of stock $i$ at time $t$; $\beta_{\rho}$ is the coefficient for each dependent variable; $Types$ is a set indicating types of COIs included in the regression. In addition to COIs, we control for explanatory variables, including daily realized volatility $\sigma_{i, t}$ and dollar volume, $vol_{i, t}$, together with six factors. The residual terms, $\epsilon_{i, t}$, are assumed to be mean zero normally distributed and the random variables $\{\epsilon_{i,t}\} $ are assumed to be independent. For inference, we apply two-tailed $t$-tests on the regression coefficients, $\beta_{\rho}$, of COIs.}

\begin{table}[htp]
    \caption{Contemporaneous Regression against Individual COIs. \\ \update{This table summarizes the coefficients of COIs by regressing again each type of COIs individually following \Cref{contemporaneous_regression_equation}. \updateSecondRound{As a benchmark, the last row shows the result of regressing against control variables only.
    } `$\beta_{\rho}$' denotes the regression coefficients and the superscript $\text{***}$ indicates significant at $1\%$ using a two-tailed t-test. `$t$' denotes the t-value of each coefficient. `adj.$R^2$' denotes the adjusted $R^{2}$ of regressions. Additional evaluation metrics are available in \Cref{tab:additional_metrics}.}} \label{tab:contemporaneous_simple_regression}
    \begin{tabularx}{\textwidth}{l*{4}{Y}}
\toprule

                      &  $\beta_{\rho}$ & t   &  adj.$R^2$ (\%)   \\ 
\midrule  

all   & 2.22*** & 34.22 & 5.66 \\
iso   & 1.70*** & 37.39 & 5.91 \\
nis   & 1.68*** & 24.78 & 4.48 \\
nis-s & 0.92*** & 16.62 & 3.44 \\
nis-c & 1.04*** & 22.76 & 3.94 \\
nis-b & 0.80*** & 18.41 & 3.50 \\

 & & & \\
control variables & & & 2.63\\

\bottomrule
\end{tabularx}

\end{table}

In \Cref{tab:contemporaneous_simple_regression}, we report results of the contemporaneous regressions against each type of COIs one at a time. Consistent with previous research, the unconditional order imbalances are positively and significantly related to returns, for almost all stocks. Furthermore, our conditional order imbalances (COIs) also express significantly positive influence on the same-day \textit{contemporaneous} returns, especially isolated COI. It is noteworthy that impacts of the three types (nis-s, nis-c and nis-b) of order imbalances derived from decomposing non-isolated trades have comparatively weaker influences with respect to their values of coefficients and t-scores. 

Focusing on the percentage of variance explained, \updateSecondRound{by comparing with the model that only uses control variables in the regression, which has an adjusted $R^2$ of 2.63\%, all types of order imbalances exhibit additional explanatory power on price impact.} Furthermore, we find that the regression with `iso' COI generates the highest adjusted $R^2$ of $5.91\%$. After decomposing trade flows, the `iso' COI, although calculated with only $28.59\%$ of trades, explains a comparable amount of variance as unconditional orde imbalance. Regressing returns against `nis' COIs achieves a lower $R^2$ than regressing against `all' or `iso' COIs. Hence, the price impact is not proportional to the quantity but appears to be driven by the types of trades. It indicates that price pressures generated by trades with distinct co-occurrence relations with other trades in the market are inhomogeneous and warrant studying separately.

In addition to the significant effect of individual conditional order imbalances on returns, we are also interested in the extra information gained from decomposing aggregated order imbalances. To this end, we fit regressions with multiple types of COIs and report the results in \Cref{tab:contemporaneous_multiple_regression}. Each regression takes as input a group of COIs as indicated in the first column. We draw inference on the coefficients. Taking the influence of feature numbers into account, we also use the adjusted $R^{2}$ as an evaluation metric. 

\begin{table}[htp]
    \caption{Contemporaneous regression against multiple COIs. \\ \update{We run regressions against multiple COIs following \Cref{contemporaneous_regression_equation}. The first column states the types of COIs input in each regression. Regression coefficients
    of different types of COIs are listed in the following columns as indicated by the column names. The superscript $\text{*, **. ***}$ indicate significant at $10\%, 5\%$ and $1\%$ respectively using a two-tailed t-test and corresponding t-values are reported in the parentheses below. The last column,`adj.$R^2$', presents the adjusted $R^{2}$ of regressions. Additional evaluation metrics are available in \Cref{tab:additional_metrics}.}}
    \label{tab:contemporaneous_multiple_regression}
    \resizebox{\textwidth}{!} {\begin{tabularx}{1.3\textwidth}{l*{8}{X}}
\toprule
                            & iso     & nis-c   & nis-b   & nis-s  & nis     & all     & adj.$R^2$ (\%) \\
\midrule

iso + nis                               & 1.42*** &          &          &          & 0.75*** &         & 6.19            \\
                                        & (28.35) &          &          &          & (10.79) &         &                 \\
iso + nis-c                             & 1.51*** & 0.50***  &          &          &         &         & 6.17            \\
                                        & (31.33) & (10.76)  &          &          &         &         &                 \\
iso + nis-c + nis-b                     & 1.51*** & 0.30***  & 0.35***  &          &         &         & 6.29            \\
                                        & (31.19) & (6.35)   & (8.85)   &          &         &         &                 \\
iso + nis-c + nis-b + nis-s             & 1.66*** & 0.21***  & 0.46***  & -0.29*** &         &         & 6.33            \\
                                        & (32.06) & (5.09)   & (12.75)  & (-5.61)  &         &         &                 \\
iso + nis-c + nis-b + nis-s + nis       & 1.67*** & -0.45*** & -0.38*** & -1.51*** & 2.74*** &         & 6.58            \\
                                        & (32.39) & (-7.47)  & (-6.32)  & (-14.24) & (13.35) &         &                 \\
iso + nis-c + nis-b + nis-s + all       & 0.88*** & -0.21*** & -0.08    & -1.06*** &         & 2.54*** & 6.51            \\
                                        & (8.82)  & (-3.93)  & (-1.43)  & (-11.10) &         & (9.70)  &                 \\
iso + nis-c + nis-b + nis-s + nis + all & 1.76*** & -0.45*** & -0.38*** & -1.53*** & 2.95*** & -0.27   & 6.59            \\
                                        & (10.97) & (-7.53)  & (-6.55)  & (-14.69) & (8.51)  & (-0.61) &       \\         

\bottomrule
\end{tabularx}}
    
\end{table}

From \Cref{tab:contemporaneous_multiple_regression}, we observe evident improvements in the adjusted $R^2$ when taking multiple trade types into account. Using the unconditional order imbalance as benchmark, splitting market orders into isolated and non-isolated explains \update{$0.53\%$ more of the total variance, \updateSecondRound{which is a 9.36\% increase from the benchmark adjusted $R^2$ of 5.66\%}. To examine the contribution of further decomposition of non-isolated trades, we add the sub-types (nis-c, nis-b, and nis-s) to \textit{iso} COI one at a time following the descending order of $R^2$ in \Cref{tab:contemporaneous_simple_regression}, and \textit{nis} COI at the end. As the adjusted $R^2$ increases, we conclude that all types of COIs of decomposed trade flows contain distinct impact on stock returns. Finally, according to the regression in the last row, in the presence of decomposed COIs, the undecomposed order imbalance is not significant for explaining price impact. In conclusion, we successfully separate trades with different contemporaneous price impact from the entire trade flow, and the decomposition helps explain contemporaneous daily price changes.}

\updateSecondRound{In conjunction with panel regression, we also perform time series regressions for individual stocks and present the results in \Cref{appendix:time_series_reg}. \Cref{tab:time_series_contemporaneous_simple_regression} suggests that the significant and positive relationships between contemporaneous returns and each type of COIs are consistent over the majority of stocks; the distributions of coefficients are illustrated in \Cref{fig:contemp_coeff_dist}. Additionally, \Cref{fig:contemp_R2_dist} sketches the density of the adjusted $R^2$ across all stocks;  
the distributions for all types of COIs are positively skewed and have mean above 10\%. }

\update{To investigate the temporal consistency, we replicate the aforementioned panel regression analysis on a yearly basis from 2017 to 2020, and report in \Cref{tab:contemporaneous_simple_regression_yearly} and \Cref{tab:contemporaneous_multiple_regression_yearly} in \Cref{appendix:subperiods}. The above conclusion remains true as above. During 2020, when the COVID-19 changed the market environment, \textit{iso} COI still has the most significant price impact, while the contemporaneous returns are less sensitive to \textit{nis-c} and \textit{nis-b} COIs. \updateSecondRound{We remark that there is a decreasing trend in adjusted $R^2$ over the years, as well as the improvement from baseline of using only the control variables.} }

\section{Predictive power of imbalances on future returns} \label{predictability}
In conjunction with contemporaneous effects of order imbalances, it is also important to study their forecasting power. In this section, we show that iso and nis-s order imbalances are positively related to future returns, while \textit{nis}, \textit{nis-c} and \textit{nis-b} COIs are negatively correlated with future returns. Moreover, we discover that decomposing trade flows and simultaneously using multiple COIs contain signals for forecasting next-day returns. We provide evidence, using both regression and portfolio sorting approaches.

\subsection{Predictive regression}
To examine the contribution of the trade flow decomposition to return forecasting, we perform the same regression analysis procedures as in the previous section. More precisely, to explore the connection between COIs and one-day ahead market-excess returns, \update{we perform panel regression on future returns, $r_{i, t+1}$, against current COIs while controlling for current returns $r_{i, t}$ as well as explanatory variables in \Cref{contemporaneous_regression_equation}, under the model 
\begin{equation} \label{predictive_regression_equation}
\begin{split}
    r_{i, t+1} = \; &\alpha + \sum_{\rho \; \in \; Types} \beta_{\rho} COI_{i, t} ^{\rho} + \beta_{\tau} r_{i, t} + \beta_{\sigma} \sigma_{i, t} + \beta_{v} vol_{i, t} \\ 
    &+ \beta_{b} MKT_{t} + \beta_{s} SMB_{t} + \beta_{h} HML_{t} + \beta_{r} RMW_{t} + \beta_{c} CMA_{t} + \beta_{m} MON_{t} + \epsilon_{i, t+1},
\end{split}\end{equation}
where $\beta_{\rho}$ is the coefficient for each dependent variable; $Types$ is a set indicating types of COIs included in the regression; and $\epsilon_{i, t}$ are the residual terms which are assumed to be independent and identically distributed with mean zero normal distributions.}

\Cref{tab:predictive_regression} documents the regression results.
As expected, unlike contemporaneous impact, both the magnitudes and percentages of significant coefficients are low, with the coefficient for unconditional order imbalances being approximately equal to zero. Over our study period, we do not find evidence to support the theoretical model put forth by \citet{chordia2004order}, which would yield a significant positive relationship between imbalances and one-day ahead returns, in the absence of future order imbalance. However, with our decomposition of trades into categories, we can strengthen the above signals. Our findings suggest that the price pressures which arose from \textit{isolated} and  \textit{non-self-isolated} order executions show moderate predictive power. Additionally, \textit{non-isolated (nis)}, \textit{non-cross-isolated} and \textit{non-both-isolated} trade imbalances are negatively associated with future price changes. Especially, \textit{iso} and \textit{nis-b} COIs exhibit significant predictive power on future returns. In term of adjusted $R^2$, all COIs of the decomposed trade flows outperform the COI of the undecomposed (i.e., aggregated) trade flow. \updateSecondRound{Additionally, the adjusted $R^2$ of regressing with only control variables is higher than incorporating unconditional order imbalance, but lower than including any types of COIs. 
This finding underscores the importance of decomposing trade flows when forecasting returns.}

\begin{table}[htp]
    \caption{Predictive regression against individual COIs. \\ \update{This table summarizes the coefficients to COIs by regressing again each type of COIs individually following \Cref{predictive_regression_equation}. \updateSecondRound{As a benchmark, the last row shows the result of regressing against control variables only.
    } `$\beta_{\rho}$' denotes the regression coefficients and the superscript $\text{***}$ indicates significant at $1\%$ using a two-tailed t-test. `$t$' denotes the t-value of each coefficient. `adj.$R^2$' denotes the adjusted $R^{2}$ of regressions. Additional evaluation metrics are available in \Cref{tab:additional_metrics}.}}
    \begin{tabularx}{\textwidth}{l*{4}{Y}}
\toprule

                      &  $\beta_{\rho}$ & t   &  adj.$R^2$ (\%)   \\ 
\midrule  

all    & 0.00     & 0.07  & 0.1013 \\
iso    & 0.08**   & 2.30  & 0.1090 \\
nis    & -0.05    & -0.85 & 0.1030 \\
nis-s  & 0.04     & 0.94  & 0.1027 \\
nis-c  & -0.05    & -1.44 & 0.1049 \\
nis-b  & -0.08**  & -1.97 & 0.1094 \\

 & & & \\
control variables & & & 0.1015\\
\bottomrule
\end{tabularx}
    \label{tab:predictive_regression}
\end{table}

In the next step, we regress future 1-day stock returns against \update{ different groups of COIs, as indicated in the first column of \Cref{tab:predictive_multiple_regression}. It is noteworthy that \textit{iso} COI shows significant predictive power in every regression setting. Although the other types of decomposed COIs do not show significance when the goal is to predict noisy daily returns, the signs of their coefficients are consistent. In addition, as the  adjusted $R^2$ grows, we find that, with the exception of \textit{nis-s} COI, all types of decompositions contribute to the return prediction task. Therefore, we conclude that the order imbalances conditioning on co-occurrence are valuable predictors for short-term return forecasting. We thus conclude that decomposing trade flows according to such COIs improves predicting future returns.}

\updateSecondRound{In addition to panel regression, we conduct time series predictive regressions for individual stocks and explain details in \Cref{appendix:time_series_reg}. \Cref{tab:time_series_predictive_simple_regression} shows that, for most of the stocks, the signs of coefficients of different types of COIs are in accord with our findings. We 
depict the distributions of coefficients in \Cref{fig:pred_coeff_dist}. Moreover, \Cref{fig:pred_R2_dist} illustrates the right-skewed distributions of adjusted $R^2$ across all stocks, corresponding to COI types.}

\update{To reinforce our findings, we perform the panel regression analysis on a yearly basis from 2017 to 2020, and report the results in \Cref{tab:predictive_regression_yearly} and \Cref{tab:predictive_multiple_regression_yearly} in \Cref{appendix:subperiods}. The signs of the relationships between future returns and different types of COIs are constant for almost all of the subperiods. Furthermore, the significance of \textit{iso} COI is persistent across all time periods, with the exception of 2020, which was an unusual market environment due to COVID-19. \updateSecondRound{In contrast to the contemporaneous price impact, the adjusted $R^2$ increase from 2017 to 2018, In addition, the adjusted $R^2$ peaks in 2020, indicating that during the tumultuous period, the market is less efficient and it takes longer for the stocks to absorb the price pressure. Therefore, the inferred COIs  exhibit forecasting power.} }

\begin{table}[htp]
    \caption{Predictive regression against multiple COIs. \\ \update{We run regressions against multiple COIs following \Cref{predictive_regression_equation}. The first column states the types of COIs input in each regression. Regression coefficients 
    for different types of COIs are listed in the following columns as indicated by the column names. The superscript $\text{*, **. ***}$ indicate significant at $10\%, 5\%$ and $1\%$ respectively using a two-tailed t-test and corresponding t-values are reported in the parentheses below. The The last column,`adj.$R^2$', presents the adjusted $R^{2}$ of regressions.}}
    \resizebox{\textwidth}{!}{\begin{tabularx}{1.3\textwidth}{l*{8}{X}}
\toprule
                            & iso     & nis-c   & nis-b   & nis-s  & nis     & all     & adj.$R^2$ (\%) \\
\midrule
iso + nis                               & 0.13*** &         &         &        & -0.14** &         & 0.1182          \\
                                        & (4.32)  &         &         &        & (-2.26) &         &                 \\
iso + nis-c                             & 0.12*** & -0.10** &         &        &         &         & 0.1185          \\
                                        & (3.00)  & (-2.31) &         &        &         &         &                 \\
iso + nis-c + nis-b                     & 0.12*** & -0.05   & -0.07*  &        &         &         & 0.1231          \\
                                        & (3.04)  & (-1.22) & (-1.79) &        &         &         &                 \\
iso + nis-c + nis-b + nis-s             & 0.11*** & -0.05   & -0.08** & 0.02   &         &         & 0.1230          \\
                                        & (3.18)  & (-1.2)  & (-2.05) & (0.56) &         &         &                 \\
iso + nis-c + nis-b + nis-s + nis       & 0.11*** & -0.03   & -0.05   & 0.05   & -0.07   &         & 0.1230          \\
                                        & (3.14)  & (-0.6)  & (-1.21) & (0.62) & (-0.45) &         &                 \\
iso + nis-c + nis-b + nis-s + all       & 0.16**  & -0.02   & -0.05   & 0.06   &         & -0.15   & 0.1234          \\
                                        & (2.41)  & (-0.48) & (-1.18) & (0.95) &         & (-0.77) &                 \\
iso + nis-c + nis-b + nis-s + nis + all & 0.21**  & -0.04   & -0.06   & 0.04   & 0.16    & -0.31   & 0.1234          \\
                                        & (2.26)  & (-0.72) & (-1.37) & (0.49) & (0.77)  & (-1.12) &      \\

\bottomrule
\end{tabularx}}
    \label{tab:predictive_multiple_regression}
\end{table}

\subsection{Imbalance-based portfolio sorting}

To bolster our findings on the positive and negative relations between future returns and different types of COI, we apply the portfolio sorting methods \citep{cattaneo2020characteristic, fama1993common} to translate order imbalances into portfolios. For each type of COI, we sort stocks according to their imbalance values, from low to high, into 5 quintile portfolios. Taking multiple features into account, we further create 5 $\times$ 5 double-sort portfolios, for every pair of COIs. The imbalance-sorted portfolios are equally weighted and have only long positions on stocks, with daily portfolio returns calculated as the average returns of all stocks in them. Backtests of imbalance-sorted portfolios, over the entire sample period from 2017-01-03 to \update{2020-12-31}, reinforce the finding that \textit{iso} and \textit{nis-s} imbalances are momentum signals, while the \textit{nis, nis-c and nis-b} imbalances are reversal signals, and that they have different influence on future returns.

\subsubsection{Single-sort portfolios}
Panel A of \Cref{tab:single_sort_portfolio} documents the annualized returns of single-sort portfolios. We note, in the first row, that the returns of the unconditional-imbalance-sorted portfolios are negative and fluctuate along quintiles, which confirms the absence of clear linear relations between unconditional order imbalance and future return. However, after performing the decomposition, we find that the growth in returns with increasing \textit{iso} order imbalance is almost monotonic, despite a slight drop in the second quintile, which reinforces its positive correlation with future returns. \update{There is also a slightly increasing trend for \textit{nis-s}, which is a sign of weak positive correlation. In contrast, we observe declines in average returns along other types of COIs, which echos our time series regression results and confirms negative correlations, altogether providing evidence for the proposed decomposition.}

Panel B shows daily COIs averaged over stocks in each portfolio. The COIs are signed, denoting that `Low' and `High' portfolios correspond to strong signals with opposite signs. We observe that the distributions of all signal strengths are roughly symmetric and centered around 0. In each row, there are no quintile portfolios consisting of stocks with indistinguishable average COI values. However, the portfolio returns are neither symmetric nor monotonic along quintiles (except \textit{iso}). By comparing returns in each row of Panel A, we observe that the magnitudes of the most positive returns are always smaller than the absolute values of the most negative returns. Therefore, we conjecture that the positive and negative impacts of COIs on future returns are asymmetric, with \update{negative impacts on future returns} being more influential. 

Furthermore, for the negative impacts, the highest magnitudes in COIs do not lead to the largest next day decreases. For example, the 'Low' and 2nd quintile portfolios of \textit{iso} COI have similar returns and, for portfolios of \update{\textit{nis-c} COI}, the 4th quintile reaches the lowest average return of \update{$-5.69\%$}, while the return of the highest quintile rise to \update{$-3.64\%$}. As interpretation of this phenomenon we propose that extreme imbalances can lead to strong reversal on the following day, because some investors aim to maintain stable levels of risk exposures. 

\begin{table}[t]
    \caption{Summary of single-sort portfolios. \\ This table shows the statistics of COI-sorted quintile portfolios. Each row contains five portfolios constructed by sorting all stocks every day by the type, indicated by its row index, of COI on the previous day from low to high and allocating each stock to the corresponding quintile portfolio indicated by the column names. The breakpoints are $20\%, 40\%, 60\% $ and $80\%$ of each type of COI calculated daily. Panel A presents the annualized return of each portfolio calculated by averaging its daily returns, from 2017-01-03 to \update{2020-12-31}, and multiplying by 252. Panel B, reports the average daily COIs of stocks included in portfolio over the sample period.}    
        \begin{tabularx}{\textwidth}{l*{6}{Y}}
        \toprule
        \multicolumn{6}{l}{\textbf{Pannel A: Annualized returns}} \\
        \midrule
                     &  Low  &  2   &  3  &  4   &  High        \\ 
                     \cmidrule(lr){2-6}  
            all   & -2.13 & -2.70 & -3.17 & -5.12 & -1.81 \\
            iso   & -4.52 & -4.72 & -4.37 & -3.57 & 2.24  \\
            nis   & 0.05  & -4.44 & -2.16 & -4.04 & -4.35 \\
            nis-s & -3.98 & -4.20 & -2.64 & -4.50 & 0.39  \\
            nis-c & 2.33  & -3.63 & -4.28 & -5.69 & -3.64 \\
            nis-b & 1.55  & -1.93 & -4.73 & -3.87 & -5.98 \\

    \end{tabularx}
    
    \begin{tabularx}{\textwidth}{l*{6}{Y}}
        \toprule
        \multicolumn{6}{l}{\textbf{Pannel B: Average daily COIs}} \\
        \midrule
                     &  Low  &  2    &  3    &  4   &  High        \\ 
                     \cmidrule(lr){2-6}  
            all   & -0.16 & -0.06 & 0.00  & 0.05 & 0.16 \\
            iso   & -0.23 & -0.09 & -0.01 & 0.07 & 0.21 \\
            nis   & -0.16 & -0.06 & 0.00  & 0.06 & 0.17 \\
            nis-s & -0.21 & -0.08 & -0.01 & 0.07 & 0.19 \\
            nis-c & -0.18 & -0.05 & 0.02  & 0.08 & 0.22 \\
            nisb  & -0.21 & -0.07 & 0.01  & 0.09 & 0.24 \\
        \bottomrule
    \end{tabularx}
    
    \label{tab:single_sort_portfolio}
\end{table}

\subsubsection{Double-sort portfolios}
To future investigate the interplay between COIs, we build portfolios by independently double-sorting on every pair of imbalances of decomposed trade flows. \Cref{tab:double_sort_portfolio} presents the annualized returns of all portfolios, where each block contains 25 portfolios by sorting on a pair of signals indicated by row and column names.

In each column of the iso$-$nis-c block, the average returns rise from low to high COIs of isolated trades. In contrast, controlled with \textit{iso} COI, the returns typically fall from low to high \textit{non-cross-isolated} COI. Double-sorting on the strongest signals generates the highest and lowest returns, on the upper-right and bottom-left corner of the block. The magnitudes of the strongest returns, \update{$18.20\%$ and $-16.62\%$}, are also amplified compared with sorting on one single signal. The same patterns and improvements appear when double-sorting on every pair of momentum and reversal COI features with \textit{iso} COI. \update{
However, the patterns for other pairs are not obvious. For example, when considering the blocks of iso$-$nis-s sorts, we do not %
observe any monotonic patterns along rows and columns. 
We conclude that \textit{iso} COI and the reversal signals carry distinct information and incorporating them simultaneously boosts predictive performance.}

\begin{sidewaystable}[htp]
    \caption{Annualized returns of double-sort portfolios. \\ This table presents annualized returns of double-sort portfolios based on every pair of COIs. Each panel contains $5 \times 5$ double-sort portfolios. To construct the portfolios, we sort all stocks every day by the two types, indicated by its row index and column name, of COIs on the previous day, from low to high, to five quintiles independently. Then intersections of the two sorts create 25 double-sort portfolios. The annualized return of each portfolio is calculated by averaging its daily returns, from 2017-01-03 to \update{2020-12-31}, and multiplying by 252.}    
    \resizebox{\textwidth}{0.3\textheight}{
        \begin{tabularx}{\textwidth}{l*{13}{Y}}
        \toprule
                 & &  Low     &    2     &     3   &    4   &   High  &  &  Low  &  2  &  3  &  4   &  High  \\ 
        \midrule
                 & &  \multicolumn{5}{c}{nis}   & &  \multicolumn{5}{c}{nis-s}  \\ 
        \cmidrule(lr){2-7} \cmidrule(lr){8-13} 

         Low     & \multirow{5}{*}{iso}    & 0.42   & -8.85 & -6.03  & -6.51  & -18.08 & \multirow{5}{*}{iso} & -3.68  & -7.31 & -0.52  & -7.20  & -5.02  \\
         2       &                         & -2.53  & -7.38 & -2.45  & -1.78  & -15.45 &                      & -7.26  & -6.62 & -2.38  & -1.49  & -5.17  \\
         3       &                         & -2.11  & -3.74 & -2.15  & -5.16  & -10.44  &                      & -6.36  & -4.11 & -4.76  & -5.08  & -1.13  \\
         4       &                         & 1.79   & -3.65 & -2.45  & -5.63  & -6.69  &                      &  2.15   & -1.67 & -5.28  & -10.61 & 1.07 \\
         High    &                         & 5.25   & 6.14  & 5.06   & -1.16  & 1.86   &                      &  -4.12  & 4.00  & 2.94   & 0.29   & 2.51  \\
         
        \cmidrule(lr){2-7} \cmidrule(lr){8-13}  
                 &  \multicolumn{5}{c}{nis-c}  & &  \multicolumn{5}{c}{nis-b}  \\
\bottomrule
    \end{tabularx}
    
   }
    \label{tab:double_sort_portfolio}
\end{sidewaystable}

\section{Economic value of conditional order imbalances} \label{economic_value}
As discussed in previous sections, there is evidence that conditional order imbalances contain signals for explaining and forecasting individual stock returns. In this section, we exploit their economic values by forming long-short portfolios using sorts. Our imbalance-based trading strategies generate conspicuous profits and significant abnormal returns. High trading profits also provide important evidence of the predictive power which the COIs of the decomposed trade flows possess.

\subsection{Long-short portfolio construction and evaluation}
We design practical trading strategies based upon imbalance-sorted quintile portfolios. At 9:30am of each trading day, we buy the first (resp., last) and short sell the last (resp., first) quintile portfolios for momentum (resp., reversal) signals with the same amount such that they are self-rebalancing. Every day, we close all position at 16:00pm to avoid overnight effects. Overall, the daily returns are the differences between the returns of the long and short imbalance-sorted portfolios.

To evaluate profitability, we compare the annualized returns of the portfolios, as well as the annualized Sharpe ratio \citep{sharpe1994sharpe}, defined as
\begin{equation} \label{sharpe_ratio_equation}
    SR_{p} := \frac{\textit{mean}(R_{p, t}) - R_{f}}{\textit{std}(R_{p,t})} \times \sqrt{252},
\end{equation}
where $R_{p, t}$ are daily returns of the portfolios and $R_{f}$ is the average daily risk-free rate, which equals $0.00625\%$ during the period of interest. 

\subsection{Profitability analysis}  \label{subsec:profitability}

\begin{table}[t]
    \caption{Profitability of long-short portfolios. \\ This table shows the annualized returns and Sharpe ratios of the long-short portfolios sorted on COIs indicated by the corresponding row indices and column names. The on- and off- diagonal values are for single- and double-sort portfolios respectively. Panel A presents the annualized return of portfolios calculated by averaging their daily returns, from 2017-01-03 to \update{2020-12-31}, and multiplying by 252. Panel B reports the annualized Sharpe ratios over the sample period calculated by \Cref{sharpe_ratio_equation}. }    
        \begin{tabularx}{\textwidth}{l*{6}{Y}}
        \toprule
        \multicolumn{6}{l}{\textbf{Panel A: Annualized returns}}          \\
        \midrule
                 &  iso  &  nis   &  nis-s  &  nis-c  &  nis-b   \\ 
         \midrule
        iso      &  6.76 &  23.33  &  0.83  &  34.87  &  26.43  \\
        nis      &       &  4.40   &  0.91   &  6.90   &  5.99    \\
        nis-s    &       &         &  4.38   &  4.57   &  2.08    \\
        nis-c    &       &         &         &  6.00   &  8.89   \\
        nis-b    &       &         &         &         &  7.50    \\
    \end{tabularx}
    
    \begin{tabularx}{\textwidth}{l*{6}{Y}}
        \toprule
        \multicolumn{6}{l}{\textbf{Panel B: Annualized Sharpe ratios}}          \\
        \midrule
                 &  iso  &  nis   &  nis-s  &  nis-c  &  nis-b  \\ 
         \midrule
        iso      &  1.20 &  1.08  &  -0.04  &  1.79   &  1.74   \\
        nis      &       &  0.48  &  -0.10  &  0.71   &  0.65   \\
        nis-s    &       &        &  0.47   &  0.37   &  0.07  \\
        nis-c    &       &        &         &  0.71   &  0.87   \\
        nis-b    &       &        &         &         &  0.97   \\
        \bottomrule
    \end{tabularx}
    \label{tab:port_sharpe}
\end{table}

We construct long-short portfolios and report their profitability measures in \Cref{tab:port_sharpe}. Panel A displays the annualized returns, with on- and off-diagonal values for single- and double-sort portfolios based on COIs indicated by row and column names. We find that incorporating multiple COIs improves the profit of portfolios, which is supporting evidence that the trade flow decomposition technique creates profitable COI signals. For example, the return of the long-short strategy corresponding to \textit{iso$-$nis} double-sort is \update{$23.33\%$}, which is $16.57\%$ and $18.93\%$ higher than simply sorting on \textit{iso} and \textit{nis} COI separately. The highest annualized return hits $34.87\%$ by double-sorting on \textit{iso} and \textit{nis-c} COIs. 
The Sharpe ratios in Panel B strengthen our findings on the economic value of COI signals. Adjusted for volatility, our trading strategies remain profitable, and double-sorting outperforms trading on signals individually. \update{The portfolio sorted on \textit{iso} and \textit{nis-c} achieves the highest Sharpe ratio of 1.79, followed by 1.74 of the \textit{iso$-$nis-b} sorted portfolio.  
Therefore, we find evidence that for investors, it is economically beneficial to incorporate multiple types of COIs when making trading decisions based on trade flow data.}

\begin{table}[t]
    \caption[LoF entry]{\update{ Abnormal returns of long-short portfolios. \\ This table documents the abnormal returns, $\alpha$, of long-short portfolios after adjusting for factors. For each long-short portfolio, we run time series regressions on portfolio excess returns against factor returns 
    \begin{equation*}
         \qquad R_{p, t} - R_{f, t} = \alpha_{p} + b_{p} MKT_{t} + s_{p} SMB_{t} + h_{p} HML_{t} + r_{p} RMW_{t} + c_{p} CMA_{t} + m_{p} MON_{t} + u_{p} UMD_{t} + e_{p, t}, 
    \end{equation*}
    where $\alpha_{p}$ is the abnormal return of the portfolio, the explanatory variables are the market, size, value, profitability, investment and momentum factors and $e_{p,t}$ is the idiosyncratic term. For inference, we apply the Newey–West estimator \citep{newey1994automatic} to correct for heteroscedasticity and auto-correlation in the residual terms. The on- and off- diagonal values are for single- and double-sort long-short portfolios respectively. The superscripts ${}^{*}$, ${}^{**}$ and ${}^{***}$ indicate statistical significance at $10\%$, $5\%$ and $1\%$, and the corresponding t-values are reported in the parentheses.}} \label{tab:port_alpha}
        \begin{tabularx}{\textwidth}{l*{6}{X}}
        \toprule
        \multicolumn{6}{l}{\textbf{Annualized alpha}}          \\
        \midrule
                 &  iso         &  nis           &  nis-s       &  nis-c              &  nis-b           \\ 
         \midrule
        iso      &  6.49***     &  20.36**      &  1.42         &  29.72***     &  25.50***        \\
                 &  (3.12)      &  (1.96)       &  (0.15)       &  (3.25)       &  (3.86)          \\
        nis      &              &  2.83         &  -0.48        &   4.23        &  4.33            \\
                 &              &  (1.08)       &  (-0.17)      &  (1.17)       &  (1.41)          \\
        nis-s    &              &               &   2.53        &   1.51        &  0.79            \\
                 &              &               &  (1.05)       &  (0.40)       &  (0.25)          \\
        nis-c    &              &               &               &  2.35         &  6.46            \\
                 &              &               &               &  (0.77)       &  (1.56)          \\
        nis-b    &              &               &               &               &  5.98**          \\
                 &              &               &               &               &  (2.14)          \\

        \bottomrule
    \end{tabularx}
\end{table}

From the perspective of asset pricing, COIs are unique and significant sources of abnormal returns. \update{To adjust for risk, we regress the excess returns of the long-short portfolios against the factors and show their alphas in \Cref{tab:port_alpha}. The factors are MKT, SMB, HML, RMW, CMA and MON. Additionally, we construct a daily rebalanced zero investment portfolio as an extra momentum factor (UMD), by sorting the returns of previous days 
in our universe of stocks and then longing the top half while shorting the bottom half. All the portfolios based on \textit{iso} COI, except \textit{iso}-\textit{nis-s}, generate statistically significant abnormal returns, providing evidence that the profits cannot be explained by common risk factors. In addition, the \textit{nis-b} COI-based single-sorted portfolio achieves significant abnormal return as well.}

\begin{figure}[t]
    \centering
    \includegraphics[width = 1\textwidth]{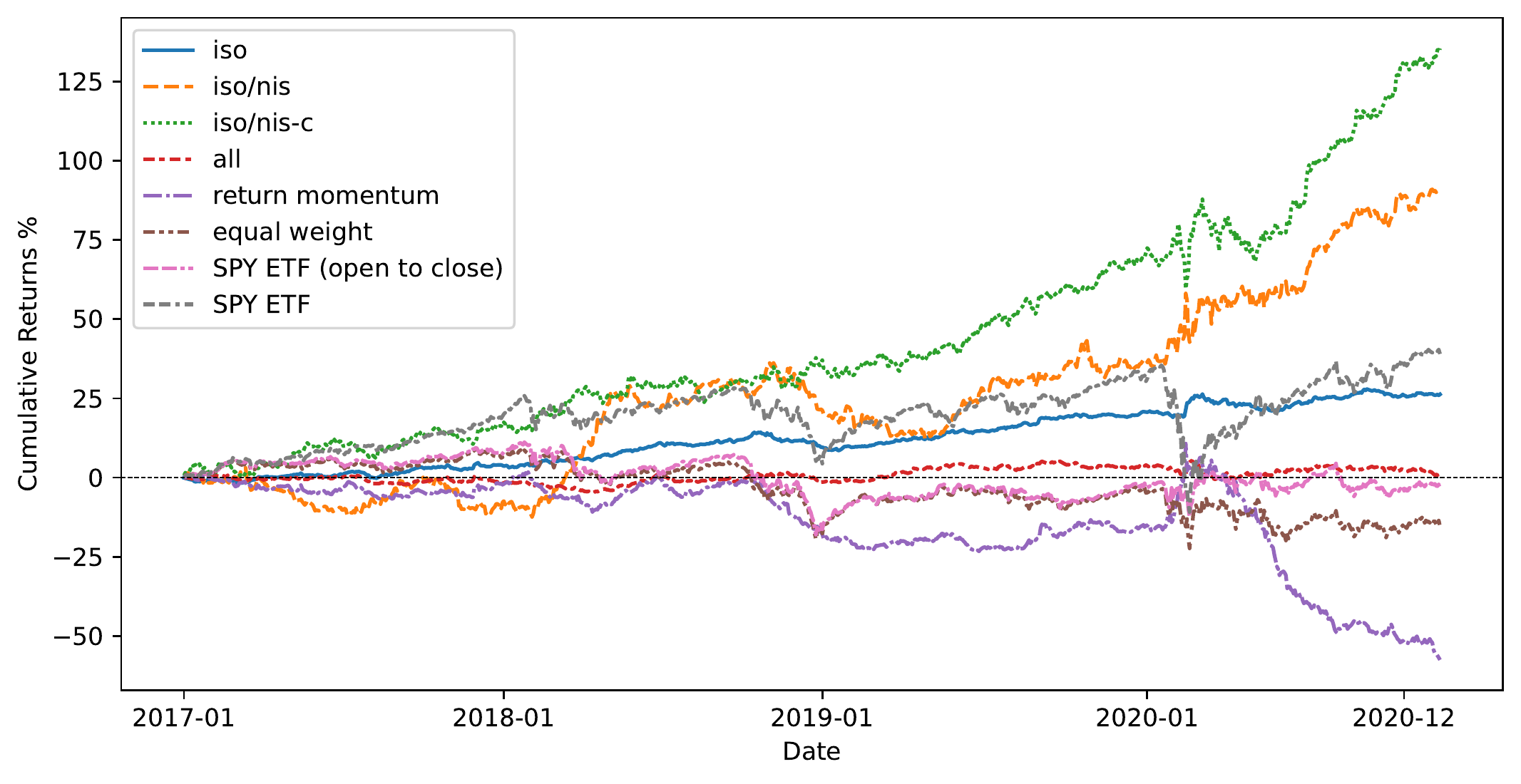}
\vspace{-6mm}
\caption{Cumulative returns of portfolios. \\ \update{ This figure plots cumulative returns of five portfolios from 2017-01-03 to 2020-12-31. The portfolios include (1) `iso': the long-short portfolio single-sorted on iso COI; (2) `iso$/$nis': the long-short portfolio double-sorted on iso and nis COIs; (3) `iso$/$nis-c': the long-short portfolio double-sorted on iso and nis-c COIs; (4) `all': the long-short portfolio single-sorted on COI of undecomposed trade flows; (5) `return momentum': the long-short portfolio single-sorted on previous day's returns; (6) `equal weight': equally-weighted portfolio of the selected 457 stocks; (7) `SPY ETF (open to close)': cumulative open to close returns of the SPDR S$\&$P 500 ETF Trust which tracks the S$\&$P 500 Index; (8) `SPY ETF': cumulative close to close returns of the SPDR S$\&$P 500 ETF.}}
\label{fig:cum_return}
\end{figure}

\begin{table}[t]
    \caption[LoF entry]{\update{ Abnormal returns of long-short portfolios. \\ This table documents the abnormal returns and relations of long-short portfolios with respect to factors. For each long-short portfolio, we run time series regressions on portfolio excess returns against factor returns 
    \begin{equation*}
         \qquad R_{p, t} - R_{f, t} = \alpha_{p} + b_{p} MKT_{t} + s_{p} SMB_{t} + h_{p} HML_{t} + r_{p} RMW_{t} + c_{p} CMA_{t} + m_{p} MON_{t} + u_{p} UMD_{t} + e_{p, t}, 
    \end{equation*}
    where $\alpha_{p}$ is the abnormal return of the portfolio, the explanatory variables are the market, size, value, profitability, investment and momentum factors, the coefficients are the exposures to the corresponding factors and $e_{p,t}$ is the idiosyncratic term. For inference, we apply the Newey–West estimator \citep{newey1994automatic} to correct for heteroscedasticity and auto-correlation in the residual terms. The superscripts ${}^{*}$, ${}^{**}$ and ${}^{***}$ indicate statistical significance at $10\%$, $5\%$ and $1\%$, and the corresponding t-values are reported in the parentheses. The last column shows the $R^2$ (unadjusted) of each regression.}} \label{tab:port_benckmark_alpha}
    \resizebox{\textwidth}{!}{ 
        \begin{tabularx}{1.3\textwidth}{l*{10}{X}}
\toprule
                        & $\alpha$    & MKT   & SMB      & HML      & RMW      & CMA      & MOM      & UMD      & $R^2$ (\%) \\
\midrule
iso             & 0.03***  & -0.01    & 0.06**  & -0.01    & 0.03    & -0.02   & -0.02   & 0.11**  & 6.24       \\
                & (3.12)   & (-0.62)  & (2.20)  & (-0.35)  & (1.16)  & (-0.44) & (-0.83) & (2.15)  &            \\
iso/nis         & 0.08**   & 0.12**   & -0.11   & -0.25*   & 0.04    & 0.45**  & -0.14   & 0.12    & 3.16       \\
                & (1.96)   & (2.17)   & (-0.85) & (-1.7)   & (0.27)  & (2.35)  & (-0.94) & (0.56)  &            \\
iso/nis-c       & 0.12***  & 0.02     & 0.26**  & -0.09    & -0.03   & 0.15    & 0.04    & -0.35** & 4.62       \\
                & (3.25)   & (0.48)   & (2.19)  & (-0.73)  & (-0.2)  & (0.88)  & (0.39) & (-2.42) &            \\
all             & 0.00     & -0.04**  & 0.15*** & 0.04*    & 0.09*** & -0.11** & 0.01    & 0.00    & 13.35      \\
\bottomrule
\end{tabularx}
    }
\end{table}

We also compare our strategies with five benchmark portfolios. Firstly, we construct a long-short portfolio of unconditional order imbalance, denoted as \textit{`all'}, to assess the economic value of trade flow decomposition. \update{Secondly, we build a return momentum benchmark portfolio constructed in the same way as COI-based long-short portfolios, but with yesterdays' market excess returns as signals. Because COIs and contemporaneous returns are significantly correlated, it is necessary to show that the profitability is not fully revealed by prices. Thirdly, we build an equally weighted portfolio with the market excess returns of the 457 stocks in our sample universe. Finally, we choose SPY, considering both open-to-close and close-to-close returns, as tradable market portfolios to benchmark against overall market performance. From the COI-based strategies, we select the single-sort portfolio of \textit{iso} COI, and the double-sort portfolios of \textit{iso}/\textit{nis} and \textit{iso}/\textit{nis-c} COIs as representatives. \Cref{fig:cum_return} visualizes cumulative returns of selected COI-based long-short portfolios and benchmarks. Over the test period, we observe that using COIs of the decomposed trade flows attains conspicuous profits. In comparison, the long-short portfolio based on undecomposed order imbalance and SPY have lower annualized returns, $0.31\%$ and $10.04\%$, and Sharpe ratios, -0.25 and 0.42 respectively. The \textit{iso} single-sort portfolio has a similar annualized return as the \update{SPY ETF}, but with much lower volatility while attaining a Sharpe ratio of \update{1.29}. In contrast, the other three benchmark portfolios, return momentum, equally weighted and SPY ETF (open-to-close), lose money over the backtest period.} Clearly, the double-sort portfolios surpasses all other portfolios with superior returns and Sharpe ratios.

\update{Furthermore, in \Cref{tab:port_benckmark_alpha}, we compare the selected portfolios abnormal returns and their relationship with other risk factors \citep{hirshleifer2010financing, chang2013pricing}. In contrast to COI-based portfolios, none of the benchmarks exhibit significant and positive abnormal returns after adjusting for the factors. In terms of factor exposures, the \textit{iso} and \textit{iso}/\textit{nis-c} portfolios have significant exposure to SMB and UMD factors, while \textit{iso}/\textit{nis-c} portfolio has significant loadings to MKT, HML and CMA factors. However, there is a large proportion of returns, for COI-based portfolios, that cannot be explained by these factors. Although the portfolios are regressed against the same set of factors, the variation explained ($R^2$) of the COI-based portfolios ranges from $3.16\%$ to $6.24\%$, attaining much lower values compared to the baseline portfolios. To be specific, the portfolio sorted on undecomposed order imbalance significantly exposes to Fama French 5 factors and has an $R^2$ of 13.35\%. Moreover, the other baseline portfolios can all be significantly explained by risk factors, with $R^2$ ranging from $55.64\%$ to $99.52\%$.}

\section{Robustness analysis} \label{robustness}
In this section, we briefly comment on the robustness of the identification of trade co-occurrences, and the construction of conditional order imbalances. Further details are provided in the Appendix.

\subsection{Neighbourhood size effect}
We replicate our analysis for eight values of $\delta$'s in \Cref{appendix:neighbourhood_size}. The patterns in contemporaneous impact and predictive power are robust for small neighbourhood sizes. Nevertheless, when $\delta$ reaches 50 milliseconds, the performance of trade co-occurrence as a filter drops. In addition, we achieve the best results of different types of COIs at different $\delta$ values, hinting at the potential benefit of the approach to combine signals derived from multiple values of $\delta$. 

\updateSecondRound{
\subsection{Representative of the market effect} \label{subsec:universe_of_stocks}
The classification of trades may depend on the choice of the representative of the market index $\mathcal{M}$. As a thought experiment, assume that 
there is a trade of Apple Inc. (AAPL) which  has only trades of Alphabet Inc. (GOOGL) in its $\delta$-neighbourhood. Then, if we replace S\&P 500 stocks with constituents of the Dow Jones Industrial Average (DOW 30) index, which include AAPL but not GOOGL, as representative of the market, the category of this AAPL trade will change from \textit{nis-c} to \textit{iso}. Therefore, we test the robustness of the trade flow decomposition by using S\&P 100 and DOW 30 respectively, and carry out comparative study. We report the details in \Cref{appendix:universe_stock}.}

\updateSecondRound{
When the set of stocks adopted as the market index, $\mathcal{M}$, varies, the fractions of each type of trades 
change slightly. 
For each type of decomposed trade flows, the COIs calculated based on different indices are highly correlated. We construct long-short portfolios corresponding to different types of COIs and universes of stocks. The portfolio double-sorted on \textit{iso} and \textit{nis-c} COIs based on S\&P 500 achieves the highest Sharpe ratios. The general results for our universe also hold when using S\&P 100 constituents for classification. However, when using Dow 30, a much smaller universe, the forecasting power deteriorates and some portfolios become nonprofitable.}

\subsection{Time-of-day effect}
Trading activities during different intraday periods have different impact on prices. As we notice, trading activities are more intensive during the first and the last half hour of each trading day. Some recent works, such as \citet{cont2021price}, exclude these volatile periods when they calculate imbalances for robustness, while others \citep{chu2021forecasting} pay special attention to imbalances during these half-hour intervals. Taking this time-of-day effect into account, we study COIs within three time intervals, namely 9:30 $-$ 10:00, 10:00 $-$ 15:30, and 15.30 $-$ 16:00 separately, and document our findings in \Cref{appendix:time_effect}. 

Our findings on contemporaneous return-imbalance relations hold for every period. Additionally, we find that the predictive power of the  decomposed trade flows originates from different time periods. The \textit{iso} and \textit{nis-s} COIs of the last hour contribute to forecasting future returns. On the other hand, the  \textit{nis-c} COI's forecasting power stems from periods other than the last half an hour. Moreover, for the  \textit{nis-b} trades, only the  COI pertaining to 10:00$-$15:30 help anticipate the next-day open-to-close market excess returns.

\subsection{COI measured by volumes}
Apart from incorporating the number of transactions, it is also common to define order imbalance as the normalized difference between volumes of buyer- and seller-initiated trades. We study the relation between individual stock returns and volume order imbalances, and analyze the corresponding trading strategies. Further details are included in \Cref{appendix:volume_coi}. 

Our findings are robust under the volume measure. We observe the same patterns as count COIs, but notice that the $R^2$ of contemporaneous regressions against volume imbalances and Sharpe ratios of corresponding long-short portfolios are generally lower than those of count imbalances, for all types of trades. This finding is in line with previous research \citep{chan1995behavior, chordia2004order} which provided evidence that the number of transactions better capture the price pressure from institutions who intend to split their orders for optimal execution.

\updateSecondRound{
\subsection{Further analysis on portfolio profitability}
To supplement the portfolio analysis in \Cref{subsec:profitability}, we further consider transaction costs for the selected and benchmark portfolios in \Cref{fig:cum_return}. We apply flat rates of round trip transaction costs, ranging from 1 to 5 basis points (bps). Our findings hold under various scenarios of costs. With rigorous backtests, the double-sorted portfolios remain profitable and outperform the benchmarks. Details are reported in \Cref{appendix:portfolio_profit}. 
}

\section{Conclusion and future directions} \label{conclusion}
In this paper, we propose the idea of trade co-occurrence, which relates trades arriving close to each other in time, and enables the study of interactions among stock transactions at a granular level. Conditional on co-occurrence with other trades, we classify every single trade into five groups. We calculate order imbalances for each type of decomposed trade flow (COI), and investigate their contemporaneous impacts and forecasting power on individual stock returns, as well as their economic value.

Our empirical results show that the decomposed trade flows have different price impacts. The COI of \textit{iso} trade flow alone can explain a comparable amount of variation in same-day returns as using COI of all trades without the  decomposition, while incorporating COIs of other trade flows further improves the explainability. For predictability, we observe that future returns, on average, are positively related with \textit{iso} and \textit{nis-s} COIs, while negatively related with \textit{nis, nis-c} and \textit{nis-b} COIs. Furthermore, the trade flow decomposition has significant economic value, and constructing long-short portfolios based on the directions of previous days' COIs leads to conspicuous enhancements in the profitability of trading strategies.

Finally, we suggest several future research directions, particularly motivated by our current limitations concerning data availability and computational power. First, we empirically show the significance of decomposing trades based on their co-occurrence with other trades, but we cannot identify who initiates certain types of trades. It would be an interesting research direction to distinguish different types of traders by leveraging private data sets \citep{tumminello2012identification, cont2021analysis}, and discover the mechanics behind the interaction of trades. For example, it would be of interest to detect whether informed traders, such as institutions, may successfully hide their trading purpose, leading to their  transactions most likely to be isolated from those of others. If high-frequency traders can be identified, it is worth applying the co-occurrence analysis to understand how HFT react to trading activities of other market participants. \updateSecondRound{Second, we have shown that the  
\mc{choice}  
of which universe of stocks is taken as the market index $\mathcal{M}$, has some influence on the decomposition. In future work, it would be 
worth investigating co-occurrence of trades within subgroups of stocks, for example industries and sectors, leading to a more fine-grained 
decomposition of the trade flows. Third, due to computational restrictions, we \mc{have used} the simple rule that if the $\delta$-neighbourhood of a trade has at least one \mc{trade}, this trade is non-isolated. Instead it could be interesting 
to consider a threshold hyperparameter when classifying trades. \mc{Furthermore, it would be interesting to investigate whether this parameter could be related to the liquidity, trading volume, and volatility of each asset.}
For example, one could use the Poisson null model to find the expected number of trades in a $\delta$-neighbourhood under complete randomness, and set a threshold value above this expectation, so that noise trades can be eliminated.}
\update{Fourth, co-occurrences of trades could be employed  to construct a pairwise similarity between stocks, which could be further leveraged to address non-synchronous trading issues, and to improve robust covariance estimation \citep{lu2023co}. Fifth, for data reduction purposes, we only study trades (i.e. the execution of limit orders against market orders), rather than all limit order book events, such as adds or cancels.} Past studies have found that submissions of new orders and cancellations of existing limit orders also lead to price impact. It would also be interesting to extend our idea to the co-occurrence of limit orders \mc{in the context of order flow imbalances} \citep{eisler2012price, cont2014price, xu2018multi, cont2021price}, 
\mc{and consider conditional order flow imbalances \citep{decomposed_OFI} analogues to our COIs}.

\bibliographystyle{rQUF}
\bibliography{rQUFguide}

\begin{thebibliography}{68}
\providecommand{\natexlab}[1]{#1}
\providecommand{\noopsort}[1]{}
\providecommand{\printfirst}[2]{#1}
\providecommand{\singleletter}[1]{#1}
\providecommand{\switchargs}[2]{#2#1}

\bibitem[\protect\citeauthoryear{Aaron
  {\itshape{et~al.}}}{2018}]{aaron2018image}
Aaron, J.S., Taylor, A.B. and Chew, T.L., Image co-localization--co-occurrence
  versus correlation. {\itshape Journal of Cell Science}, 2018, \textbf{131},
  jcs211847.

\bibitem[\protect\citeauthoryear{A{\"\i}t-Sahalia
  {\itshape{et~al.}}}{2022}]{ait2022and}
A{\"\i}t-Sahalia, Y., Fan, J., Xue, L. and Zhou, Y., How and When are
  High-Frequency Stock Returns Predictable?. Technical report, National Bureau
  of Economic Research, 2022.

\bibitem[\protect\citeauthoryear{Aldridge}{2013}]{aldridge2013high}
Aldridge, I., {\itshape High-frequency trading: a practical guide to
  algorithmic strategies and trading systems},  Vol. 604, , 2013, John Wiley \&
  Sons.

\bibitem[\protect\citeauthoryear{Appel and Holden}{1998}]{appel1998co}
Appel, A.E. and Holden, G.W., The co-occurrence of spouse and physical child
  abuse: a review and appraisal.. {\itshape Journal of Family Psychology},
  1998, \textbf{12}, 578.

\bibitem[\protect\citeauthoryear{Ara{\'u}jo
  {\itshape{et~al.}}}{2011}]{araujo2011using}
Ara{\'u}jo, M.B., Rozenfeld, A., Rahbek, C. and Marquet, P.A., Using species
  co-occurrence networks to assess the impacts of climate change. {\itshape
  Ecography}, 2011, \textbf{34}, 897--908.

\bibitem[\protect\citeauthoryear{Bailey
  {\itshape{et~al.}}}{2009}]{bailey2009stock}
Bailey, W., Cai, J., Cheung, Y.L. and Wang, F., Stock returns, order
  imbalances, and commonality: Evidence on individual, institutional, and
  proprietary investors in China. {\itshape Journal of Banking \& Finance},
  2009, \textbf{33}, 9--19.

\bibitem[\protect\citeauthoryear{Bechler and
  Ludkovski}{2015}]{bechler2015optimal}
Bechler, K. and Ludkovski, M., Optimal execution with dynamic order flow
  imbalance. {\itshape SIAM Journal on Financial Mathematics}, 2015,
  \textbf{6}, 1123--1151.

\bibitem[\protect\citeauthoryear{Brunnermeier and
  Pedersen}{2005}]{brunnermeier2005predatory}
Brunnermeier, M.K. and Pedersen, L.H., Predatory trading. {\itshape The Journal
  of Finance}, 2005, \textbf{60}, 1825--1863.

\bibitem[\protect\citeauthoryear{Carhart}{1997}]{carhart1997persistence}
Carhart, M.M., On persistence in mutual fund performance. {\itshape The Journal
  of Finance}, 1997, \textbf{52}, 57--82.

\bibitem[\protect\citeauthoryear{Cartea
  {\itshape{et~al.}}}{2015}]{cartea2015algorithmic}
Cartea, {\'A}., Jaimungal, S. and Penalva, J., {\itshape Algorithmic and
  high-frequency trading}, 2015, Cambridge University Press.

\bibitem[\protect\citeauthoryear{Cattaneo
  {\itshape{et~al.}}}{2020}]{cattaneo2020characteristic}
Cattaneo, M.D., Crump, R.K., Farrell, M.H. and Schaumburg, E.,
  Characteristic-sorted portfolios: Estimation and inference. {\itshape Review
  of Economics and Statistics}, 2020, \textbf{102}, 531--551.

\bibitem[\protect\citeauthoryear{Chakravarty
  {\itshape{et~al.}}}{2012}]{chakravarty2012clean}
Chakravarty, S., Jain, P., Upson, J. and Wood, R., Clean sweep: Informed
  trading through intermarket sweep orders. {\itshape Journal of Financial and
  Quantitative Analysis}, 2012, \textbf{47}, 415--435.

\bibitem[\protect\citeauthoryear{Chan and Lakonishok}{1995}]{chan1995behavior}
Chan, L.K. and Lakonishok, J., The behavior of stock prices around
  institutional trades. {\itshape The Journal of Finance}, 1995, \textbf{50},
  1147--1174.

\bibitem[\protect\citeauthoryear{Chang}{2012}]{chang2012order}
Chang, C.Y., Order imbalance and daily momentum investing: Evidence from
  Taiwan. {\itshape Financial Review}, 2012, \textbf{47}, 697--718.

\bibitem[\protect\citeauthoryear{Chang
  {\itshape{et~al.}}}{2013}]{chang2013pricing}
Chang, E.C., Luo, Y. and Ren, J., Pricing deviation, misvaluation comovement,
  and macroeconomic conditions. {\itshape Journal of Banking \& Finance}, 2013,
  \textbf{37}, 5285--5299.

\bibitem[\protect\citeauthoryear{Chordia
  {\itshape{et~al.}}}{2016}]{chordia2016buyers}
Chordia, T., Goyal, A. and Jegadeesh, N., Buyers versus sellers: who initiates
  trades, and when?. {\itshape Journal of Financial and Quantitative Analysis},
  2016, \textbf{51}, 1467--1490.

\bibitem[\protect\citeauthoryear{Chordia
  {\itshape{et~al.}}}{2002}]{chordia2002order}
Chordia, T., Roll, R. and Subrahmanyam, A., Order imbalance, liquidity, and
  market returns. {\itshape Journal of Financial Economics}, 2002, \textbf{65},
  111--130.

\bibitem[\protect\citeauthoryear{Chordia and
  Subrahmanyam}{2004}]{chordia2004order}
Chordia, T. and Subrahmanyam, A., Order imbalance and individual stock returns:
  Theory and evidence. {\itshape Journal of Financial Economics}, 2004,
  \textbf{72}, 485--518.

\bibitem[\protect\citeauthoryear{Chu and Qiu}{2021}]{chu2021forecasting}
Chu, X. and Qiu, J., Forecasting stock returns using first half an hour order
  imbalance. {\itshape International Journal of Finance \& Economics}, 2021,
  \textbf{26}, 3236--3245.

\bibitem[\protect\citeauthoryear{Cont
  {\itshape{et~al.}}}{2021{\natexlab{a}}}]{cont2021analysis}
Cont, R., Cucuringu, M., Glukhov, V. and Prenzel, F., Analysis and modeling of
  client order flow in limit order markets. {\itshape Available at SSRN},
  2021{\natexlab{a}}.

\bibitem[\protect\citeauthoryear{Cont
  {\itshape{et~al.}}}{2021{\natexlab{b}}}]{cont2021price}
Cont, R., Cucuringu, M. and Zhang, C., Price impact of order flow imbalance:
  multi-level, cross-sectional and forecasting. {\itshape arXiv e-prints},
  2021{\natexlab{b}}, pp. arXiv--2112.

\bibitem[\protect\citeauthoryear{Cont {\itshape{et~al.}}}{2014}]{cont2014price}
Cont, R., Kukanov, A. and Stoikov, S., The price impact of order book events.
  {\itshape Journal of Financial Econometrics}, 2014, \textbf{12}, 47--88.

\bibitem[\protect\citeauthoryear{Cox}{2021}]{cox2021iso}
Cox, J., ISO order imbalances and individual stock returns. {\itshape Journal
  of Financial Research}, 2021, \textbf{44}, 5--23.

\bibitem[\protect\citeauthoryear{Dagan
  {\itshape{et~al.}}}{1999}]{dagan1999similarity}
Dagan, I., Lee, L. and Pereira, F.C., Similarity-based models of word
  cooccurrence probabilities. {\itshape Machine Learning}, 1999, \textbf{34},
  43--69.

\bibitem[\protect\citeauthoryear{Donges
  {\itshape{et~al.}}}{2016}]{donges2016event}
Donges, J.F., Schleussner, C.F., Siegmund, J.F. and Donner, R.V., Event
  coincidence analysis for quantifying statistical interrelationships between
  event time series. {\itshape The European Physical Journal Special Topics},
  2016, \textbf{225}, 471--487.

\bibitem[\protect\citeauthoryear{Eisler
  {\itshape{et~al.}}}{2012}]{eisler2012price}
Eisler, Z., Bouchaud, J.P. and Kockelkoren, J., The price impact of order book
  events: market orders, limit orders and cancellations. {\itshape Quantitative
  Finance}, 2012, \textbf{12}, 1395--1419.

\bibitem[\protect\citeauthoryear{Fama and French}{1992}]{fama1992cross}
Fama, E.F. and French, K.R., The Cross-Section of Expected Stock Returns.
  {\itshape The Journal of Finance}, 1992, \textbf{47}, 427--465.

\bibitem[\protect\citeauthoryear{Fama and French}{1993}]{fama1993common}
Fama, E.F. and French, K.R., Common risk factors in the returns on stocks and
  bonds. {\itshape Journal of Financial Economics}, 1993, \textbf{33}, 3--56.

\bibitem[\protect\citeauthoryear{Fama and French}{2015}]{fama2015five}
Fama, E.F. and French, K.R., A five-factor asset pricing model. {\itshape
  Journal of Financial Economics}, 2015, \textbf{116}, 1--22.

\bibitem[\protect\citeauthoryear{Foster and
  Viswanathan}{1996}]{foster1996strategic}
Foster, F.D. and Viswanathan, S., Strategic trading when agents forecast the
  forecasts of others. {\itshape The Journal of Finance}, 1996, \textbf{51},
  1437--1478.

\bibitem[\protect\citeauthoryear{Galleguillos
  {\itshape{et~al.}}}{2008}]{galleguillos2008object}
Galleguillos, C., Rabinovich, A. and Belongie, S., Object categorization using
  co-occurrence, location and appearance. In {\itshape Proceedings of the
  }{\itshape 2008 IEEE Conference on Computer Vision and Pattern Recognition},
  pp. 1--8, 2008.

\bibitem[\protect\citeauthoryear{Gotelli}{2000}]{gotelli2000null}
Gotelli, N.J., Null model analysis of species co-occurrence patterns. {\itshape
  Ecology}, 2000, \textbf{81}, 2606--2621.

\bibitem[\protect\citeauthoryear{Grossman and
  Miller}{1988}]{grossman1988liquidity}
Grossman, S.J. and Miller, M.H., Liquidity and market structure. {\itshape the
  Journal of Finance}, 1988, \textbf{43}, 617--633.

\bibitem[\protect\citeauthoryear{Guilbaud and Pham}{2013}]{guilbaud2013optimal}
Guilbaud, F. and Pham, H., Optimal high-frequency trading with limit and market
  orders. {\itshape Quantitative Finance}, 2013, \textbf{13}, 79--94.

\bibitem[\protect\citeauthoryear{Guo {\itshape{et~al.}}}{2017}]{guo2017news}
Guo, L., Peng, L., Tao, Y. and Tu, J., News co-occurrence, attention spillover,
  and return predictability. {\itshape arXiv preprint arXiv:1703.02715}, 2017.

\bibitem[\protect\citeauthoryear{Hagstr{\"o}mer and
  Nord{\'e}n}{2013}]{hagstromer2013diversity}
Hagstr{\"o}mer, B. and Nord{\'e}n, L., The diversity of high-frequency traders.
  {\itshape Journal of Financial Markets}, 2013, \textbf{16}, 741--770.

\bibitem[\protect\citeauthoryear{Hirschey}{2021}]{hirschey2021high}
Hirschey, N., Do high-frequency traders anticipate buying and selling
  pressure?. {\itshape Management Science}, 2021, \textbf{67}, 3321--3345.

\bibitem[\protect\citeauthoryear{Hirshleifer and
  Jiang}{2010}]{hirshleifer2010financing}
Hirshleifer, D. and Jiang, D., A financing-based misvaluation factor and the
  cross-section of expected returns. {\itshape The Review of Financial
  Studies}, 2010, \textbf{23}, 3401--3436.

\bibitem[\protect\citeauthoryear{Huang and Polak}{2011}]{huang2011lobster}
Huang, R. and Polak, T., Lobster: Limit order book reconstruction system.
  {\itshape Available at SSRN 1977207}, 2011.

\bibitem[\protect\citeauthoryear{Jegadeesh and
  Titman}{1993}]{jegadeesh1993returns}
Jegadeesh, N. and Titman, S., Returns to buying winners and selling losers:
  Implications for stock market efficiency. {\itshape The Journal of Finance},
  1993, \textbf{48}, 65--91.

\bibitem[\protect\citeauthoryear{Kolesnikova}{2016}]{kolesnikova2016survey}
Kolesnikova, O., Survey of word co-occurrence measures for collocation
  detection. {\itshape Computaci{\'o}n y Sistemas}, 2016, \textbf{20},
  327--344.

\bibitem[\protect\citeauthoryear{Kolm {\itshape{et~al.}}}{2021}]{kolm2021deep}
Kolm, P.N., Turiel, J. and Westray, N., Deep order flow imbalance: Extracting
  alpha at multiple horizons from the limit order book. {\itshape Available at
  SSRN 3900141}, 2021.

\bibitem[\protect\citeauthoryear{Kraus and Stoll}{1972}]{kraus1972parallel}
Kraus, A. and Stoll, H.R., Parallel trading by institutional investors.
  {\itshape Journal of Financial and Quantitative Analysis}, 1972, \textbf{7},
  2107--2138.

\bibitem[\protect\citeauthoryear{Kyle}{1985}]{kyle1985continuous}
Kyle, A.S., Continuous auctions and insider trading. {\itshape Econometrica:
  Journal of the Econometric Society}, 1985, pp. 1315--1335.

\bibitem[\protect\citeauthoryear{Kyle {\itshape{et~al.}}}{2011}]{kyle2011model}
Kyle, A.S., Ou-Yang, H. and Wei, B., A model of portfolio delegation and
  strategic trading. {\itshape The Review of Financial Studies}, 2011,
  \textbf{24}, 3778--3812.

\bibitem[\protect\citeauthoryear{Lee {\itshape{et~al.}}}{2004}]{lee2004order}
Lee, Y.T., Liu, Y.J., Roll, R. and Subrahmanyam, A., Order imbalances and
  market efficiency: Evidence from the Taiwan Stock Exchange. {\itshape Journal
  of Financial and Quantitative Analysis}, 2004, \textbf{39}, 327--341.

\bibitem[\protect\citeauthoryear{Lu {\itshape{et~al.}}}{2023}]{lu2023co}
Lu, Y., Reinert, G. and Cucuringu, M., Co-trading networks for modeling dynamic
  interdependency structures and estimating high-dimensional covariances in US
  equity markets. {\itshape arXiv preprint arXiv:2302.09382}, 2023.

\bibitem[\protect\citeauthoryear{Lucchese
  {\itshape{et~al.}}}{2022}]{lucchese2022short}
Lucchese, L., Pakkanen, M. and Veraart, A., The Short-Term Predictability of
  Returns in Order Book Markets: a Deep Learning Perspective. {\itshape arXiv
  preprint arXiv:2211.13777}, 2022.

\bibitem[\protect\citeauthoryear{Ma {\itshape{et~al.}}}{2011}]{ma2011mining}
Ma, Z., Pant, G. and Sheng, O.R., Mining competitor relationships from online
  news: A network-based approach. {\itshape Electronic Commerce Research and
  Applications}, 2011, \textbf{10}, 418--427.

\bibitem[\protect\citeauthoryear{MacKenzie
  {\itshape{et~al.}}}{2004}]{mackenzie2004investigating}
MacKenzie, D.I., Bailey, L.L. and Nichols, J.D., Investigating species
  co-occurrence patterns when species are detected imperfectly. {\itshape
  Journal of Animal Ecology}, 2004, \textbf{73}, 546--555.

\bibitem[\protect\citeauthoryear{Newey and West}{1994}]{newey1994automatic}
Newey, W.K. and West, K.D., Automatic lag selection in covariance matrix
  estimation. {\itshape The Review of Economic Studies}, 1994, \textbf{61},
  631--653.

\bibitem[\protect\citeauthoryear{O’Hara}{2015}]{o2015high}
O’Hara, M., High frequency market microstructure. {\itshape Journal of
  Financial Economics}, 2015, \textbf{116}, 257--270.

\bibitem[\protect\citeauthoryear{Scharfstein and
  Stein}{1990}]{scharfstein1990herd}
Scharfstein, D.S. and Stein, J.C., Herd behavior and investment. {\itshape The
  American Economic Review}, 1990, pp. 465--479.

\bibitem[\protect\citeauthoryear{Sharpe}{1994}]{sharpe1994sharpe}
Sharpe, W.F., The sharpe ratio. {\itshape Journal of Portfolio Management},
  1994, \textbf{21}, 49--58.

\bibitem[\protect\citeauthoryear{Shenoy and Zhang}{2007}]{shenoy2007order}
Shenoy, C. and Zhang, Y.J., Order imbalance and stock returns: Evidence from
  China. {\itshape The Quarterly Review of Economics and Finance}, 2007,
  \textbf{47}, 637--650.

\bibitem[\protect\citeauthoryear{Sitaru
  {\itshape{et~al.}}}{2023}]{decomposed_OFI}
Sitaru, B., Calinescu, A. and Cucuringu, M., {Order Flow Decomposition for
  Price Impact Analysis in Equity Limit Order Books}. {\itshape to appear in
  Proceedings of the Fourth ACM International Conference on AI in Finance
  (ICAIF 2023); SSRN:4572510}, 2023.

\bibitem[\protect\citeauthoryear{Spiegel and
  Subrahmanyam}{1995}]{spiegel1995intraday}
Spiegel, M. and Subrahmanyam, A., On intraday risk premia. {\itshape The
  Journal of Finance}, 1995, \textbf{50}, 319--339.

\bibitem[\protect\citeauthoryear{Stoll}{1978}]{stoll1978supply}
Stoll, H.R., The supply of dealer services in securities markets. {\itshape The
  Journal of Finance}, 1978, \textbf{33}, 1133--1151.

\bibitem[\protect\citeauthoryear{Tang {\itshape{et~al.}}}{2019}]{tang2019news}
Tang, Y., Zhou, Y. and Hong, M., News co-occurrences, stock return
  correlations, and portfolio construction implications. {\itshape Journal of
  Risk and Financial Management}, 2019, \textbf{12}, 45.

\bibitem[\protect\citeauthoryear{Tumminello
  {\itshape{et~al.}}}{2012}]{tumminello2012identification}
Tumminello, M., Lillo, F., Piilo, J. and Mantegna, R.N., Identification of
  clusters of investors from their real trading activity in a financial market.
  {\itshape New Journal of Physics}, 2012, \textbf{14}, 013041.

\bibitem[\protect\citeauthoryear{Van~Kervel and Menkveld}{2019}]{van2019high}
Van~Kervel, V. and Menkveld, A.J., High-frequency trading around large
  institutional orders. {\itshape The Journal of Finance}, 2019, \textbf{74},
  1091--1137.

\bibitem[\protect\citeauthoryear{Wang {\itshape{et~al.}}}{2021}]{wang2021high}
Wang, Q., Teng, B., Hao, Q. and Shi, Y., High-frequency statistical arbitrage
  strategy based on stationarized order flow imbalance. {\itshape Procedia
  Computer Science}, 2021, \textbf{187}, 518--523.

\bibitem[\protect\citeauthoryear{Wu {\itshape{et~al.}}}{2019}]{wu2019deep}
Wu, Q., Zhang, Z., Pizzoferroto, A., Cucuringu, M. and Liu, Z., A deep learning
  framework for pricing financial instruments. {\itshape arXiv.org}, 2019.

\bibitem[\protect\citeauthoryear{Xu {\itshape{et~al.}}}{2018}]{xu2018multi}
Xu, K., Gould, M.D. and Howison, S.D., Multi-level order-flow imbalance in a
  limit order book. {\itshape Market Microstructure and Liquidity}, 2018,
  \textbf{4}, 1950011.

\bibitem[\protect\citeauthoryear{Yang and Zhu}{2020}]{yang2020back}
Yang, L. and Zhu, H., Back-running: Seeking and hiding fundamental information
  in order flows. {\itshape The Review of Financial Studies}, 2020,
  \textbf{33}, 1484--1533.

\bibitem[\protect\citeauthoryear{Ye {\itshape{et~al.}}}{2017}]{ye2017co}
Ye, S., Zeng, G., Wu, H., Zhang, C., Liang, J., Dai, J., Liu, Z., Xiong, W.,
  Wan, J., Xu, P. {\itshape et~al.}, Co-occurrence and interactions of
  pollutants, and their impacts on soil remediation—a review. {\itshape
  Critical Reviews in Environmental Science and Technology}, 2017, \textbf{47},
  1528--1553.

\bibitem[\protect\citeauthoryear{Zhang
  {\itshape{et~al.}}}{2019{\natexlab{a}}}]{zhang2019order}
Zhang, T., Gu, G.F. and Zhou, W.X., Order imbalances and market efficiency: New
  evidence from the Chinese stock market. {\itshape Emerging Markets Review},
  2019{\natexlab{a}}, \textbf{38}, 458--467.

\bibitem[\protect\citeauthoryear{Zhang
  {\itshape{et~al.}}}{2019{\natexlab{b}}}]{zhang2019deeplob}
Zhang, Z., Zohren, S. and Roberts, S., Deeplob: Deep convolutional neural
  networks for limit order books. {\itshape IEEE Transactions on Signal
  Processing}, 2019{\natexlab{b}}, \textbf{67}, 3001--3012.

\end{thebibliography}

\appendices

\section{Sample universe of stocks}  \label{appendix:sampe_stocks}

\Cref{tab:sum_stats_sample_stocks} provides a brief summary of the number of stocks we use in this study. 

\begin{table}[h]
    \caption{Description of the sample universe. \\ This table summarize the total number of stocks, as well as, the number of stocks grouped by their sector membership.} 
        \begin{tabularx}{\textwidth}{l*{2}{Y}}
        \toprule
         &   Number of stocks   \\    
         \midrule
    Total    &    457 \\
    \\
    \textbf{\textit{Sectors:}} &  \\
    
    Communication srvices    &  24  \\
    Consumer discretionary   &  57  \\
    Consumer staples        &  32   \\
    Energy      &   24   \\ 
    Financials        &   62  \\
    Health care  &   55                    \\
    Industrials      &    60   \\
    Information technology   &     64   \\
    Materials       &    24          \\
    Real estate     &    29          \\
    Utilities       &    26          \\
    \bottomrule
    \end{tabularx}
    \label{tab:sum_stats_sample_stocks}
\end{table}

\newpage
\section{Regression analysis of subperiods} \label{appendix:subperiods}
\update{
To supplement the results in \Cref{contamporaneous_analysis} and \Cref{predictability}, we perform the regression analysis in the same settings, on a yearly basis. \Cref{tab:contemporaneous_simple_regression_yearly} and \Cref{tab:contemporaneous_multiple_regression_yearly} show the results for contemporaneous regressions, and \Cref{tab:predictive_regression_yearly} and \Cref{tab:predictive_multiple_regression_yearly} document predictive regressions. }

\begin{table}[htp]
    \caption{\update{Yearly contemporaneous regression against individual COIs. \\ This table summarizes the coefficients for COIs by regressing again each type of COIs individually following \Cref{contemporaneous_regression_equation} on a yearly basis for 2017, 2018, 2019 and 2020 respectively. \updateSecondRound{As a benchmark, 
    the last row of each panel shows the result of regressing against control variables only.} 
    `$\beta_{\rho}$' denotes the regression coefficients and the superscript $\text{***}$ indicates significant at $1\%$ using a two-tailed t-test. `$t$' denotes the t-value of each coefficient. `adj.$R^2$' denotes the adjusted $R^{2}$ of regressions.}} \label{tab:contemporaneous_simple_regression_yearly}
    \begin{tabularx}{\textwidth}{l*{4}{Y}}
\toprule
2017          & $\beta_{\rho}$ & t   &  adj.$R^2$ (\%)  \\ 
\midrule
all           & 2.43***     & 32.94   & 8.39            \\
iso           & 1.96***     & 33.48   & 9.12           \\
nis           & 1.72***     & 26.32   & 4.94           \\
nis-s         & 1.10***     & 22.55   & 3.12             \\
nis-c         & 0.89***     & 17.87   & 2.79            \\
nis-b         & 0.66***     & 16.02   & 2.08             \\
control variables & & & 0.52 \\
\bottomrule

\end{tabularx}
\end{table}

\begin{table}[htp]
    \caption{\update{Yearly contemporaneous regression against multiple COIs. \\ We run regressions against multiple COIs following \Cref{contemporaneous_regression_equation} on a yearly basis for 2017, 2018, 2019 and 2020 respectively. The first column states the types of COIs input in each regression. Regression coefficients to different types of COIs are listed in the following columns as indicated by the column names. The superscript $\text{*, **. ***}$ indicate significant at $10\%, 5\%$ and $1\%$ respectively using a two-tailed t-test and corresponding t-values are reported in the parentheses below. The The last column,`adj.$R^2$', presents the adjusted $R^{2}$ of regressions.}}
    \resizebox{\textwidth}{0.48\textheight} {\begin{tabularx}{1.3\textwidth}{l*{8}{X}}
 
\toprule
2017                                        & iso     & nis-c    & nis-b   & nis-s  & nis      & all     & adj.$R^2$ (\%) \\
\midrule

iso + nis                               & 1.68*** &          &          &          & 0.72*** &         & 9.70            \\
                                        & (28.47) &          &          &          & (11.71) &         &                 \\
iso + nis-c                             & 1.83*** & 0.39***  &          &          &         &         & 9.51            \\
                                        & (32.12) & (8.67)   &          &          &         &         &                 \\
iso + nis-c + nis-b                     & 1.83*** & 0.18***  & 0.33***  &          &         &         & 9.78            \\
                                        & (32.17) & (4.77)   & (10.17)  &          &         &         &                 \\
iso + nis-c + nis-b + nis-s             & 1.88*** & 0.16***  & 0.35***  & -0.09**  &         &         & 9.79            \\
\bottomrule
\end{tabularx}}
    \label{tab:contemporaneous_multiple_regression_yearly}
\end{table}

\begin{table}[htp]
    \caption{\update{Yearly predictive regression against individual COIs. \\ This table summarizes the coefficients for COIs by regressing again each type of COIs individually following \Cref{predictive_regression_equation} on a yearly basis for 2017, 2018, 2019 and 2020 respectively. \updateSecondRound{As a benchmark, 
    the last row of each panel shows the result of regressing against control variables only.}`$\beta_{\rho}$' denotes the regression coefficients and the superscript $\text{***}$ indicates significant at $1\%$ using a two-tailed t-test. `$t$' denotes the t-value of each coefficient. `adj.$R^2$' denotes the adjusted $R^{2}$ of regressions.}}
    \resizebox{\textwidth}{!}{
        \begin{tabularx}{\textwidth}{l*{4}{Y}}
\toprule

2017                      &  $\beta_{\rho}$ & t   &  adj.$R^2$ (\%)   \\ 
\midrule  

all   & -0.01    & -0.26 & 0.0273 \\
iso   & 0.07**   & 2.47  & 0.0360 \\
nis   & -0.06*   & -1.69 & 0.0320 \\
nis-s & 0.05**   & 2.18  & 0.0331 \\
nis-c & -0.10*** & -3.03 & 0.0569 \\
nis-b & -0.07*** & -2.75 & 0.0458 \\
control variables & & & 0.0281 \\

\toprule
2018                      &  $\beta_{\rho}$ & t   &  adj.$R^2$ (\%)   \\ 
\midrule  

all   & 0.04     & 0.73  & 0.1248 \\
iso   & 0.09**   & 2.34  & 0.1361 \\
nis   & -0.02    & -0.46 & 0.1242 \\
nis-s & 0.06     & 1.48  & 0.1285 \\
nis-c & -0.06    & -1.21 & 0.1292 \\
nis-b & -0.05    & -1.20 & 0.1275 \\
control variables & & & 0.1246 \\

\toprule
2019     &  $\beta_{\rho}$ & t   &  adj.$R^2$ (\%)   \\ 
\midrule  
all   & 0.04     & 0.91  & 0.0843 \\
iso   & 0.09***  & 2.90  & 0.0979 \\
nis   & 0.00     & -0.10 & 0.0828 \\
nis-s & 0.11***  & 3.16  & 0.0993 \\
nis-c & -0.08**  & -2.13 & 0.0953 \\
nis-b & -0.07*   & -1.79 & 0.0910 \\
control variables & & & 0.0837 \\

\toprule
2020    &  $\beta_{\rho}$ & t   &  adj.$R^2$ (\%)   \\ 
\midrule  
all   & -0.07    & -0.47 & 0.2804 \\
iso   & 0.08     & 1.05  & 0.2820 \\
nis   & -0.15    & -0.89 & 0.2850 \\
nis-s & -0.06    & -0.52 & 0.2806 \\
nis-c & 0.00     & -0.04 & 0.2791 \\
nis-b & -0.17    & -1.25 & 0.2927 \\
control variables & & & 0.2799 \\

\bottomrule
\end{tabularx}
    }
    \label{tab:predictive_regression_yearly}
\end{table}

\begin{table}[htp]
    \caption{\update{Yearly predictive regression against multiple COIs. \\ We run regressions against multiple COIs following \Cref{predictive_regression_equation} on a yearly basis for 2017, 2018, 2019 and 2020 respectively. The first column states the types of COIs input in each regression. Regression coefficients to different types of COIs are listed in the following columns as indicated by the column names. The superscript $\text{*, **. ***}$ indicate significant at $10\%, 5\%$ and $1\%$ respectively using a two-tailed t-test and corresponding t-values are reported in the parentheses below. The The last column,`adj.$R^2$', presents the adjusted $R^{2}$ of regressions.}} 
    \label{tab:predictive_multiple_regression_yearly}
    \resizebox{\textwidth}{0.48\textheight} {\begin{tabularx}{1.3\textwidth}{l*{8}{X}}

\toprule
2017                                        & iso     & nis-c    & nis-b   & nis-s  & nis      & all     & adj.$R^2$ (\%) \\
\midrule

iso + nis                               & 0.11*** &          &         &        & -0.12*** &         & 0.0511          \\
                                        & (3.28)  &          &         &        & (-2.75)  &         &                 \\
iso + nis-c                             & 0.11*** & -0.13*** &         &        &          &         & 0.0780          \\
                                        & (3.53)  & (-3.49)  &         &        &          &         &                 \\
iso + nis-c + nis-b                     & 0.11*** & -0.11*** & -0.03   &        &          &         & 0.0794          \\
                                        & (3.52)  & (-3.19)  & (-1.33) &        &          &         &                 \\
iso + nis-c + nis-b + nis-s             & 0.09*** & -0.11*** & -0.04   & 0.02   &          &         & 0.0792          \\
\bottomrule
\end{tabularx}}
\end{table}

\newpage
\section{Time series regression and distribution of $\beta$}  \label{appendix:time_series_reg}

\updateSecondRound{
In this section, we conduct contemporaneous (\Cref{contemporaneous_regression_equation}) and predictive (\Cref{predictive_regression_equation}) regressions against each type of COI, on each stock individually, instead of the panel regressions reported in \Cref{contamporaneous_analysis} and \Cref{predictability}. 
}

\subsection{Contemporaneous Time Series Regression}

\begin{figure}[t]
    \centering
    \includegraphics[width = \textwidth]{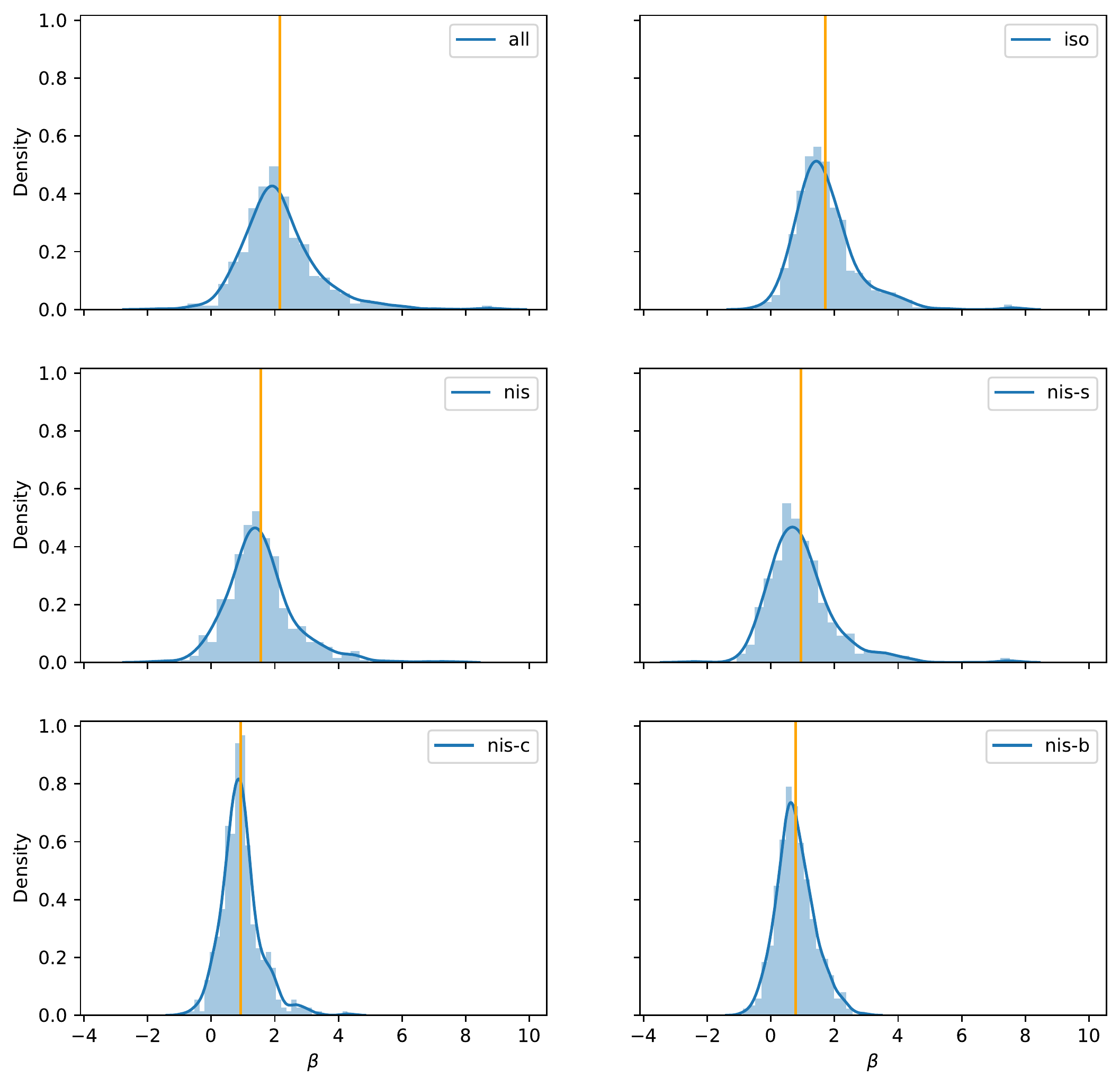}
    \caption{\updateSecondRound{Distributions of coefficients of contemporaneous time series regressions. \\ This table displays the histogram and kernel density estimation of the coefficients of contemporaneous time series regressions. The orange line indicates the mean of the coefficients.}}
    \label{fig:contemp_coeff_dist}
\end{figure}

\begin{figure}[t]
    \centering
    \includegraphics[width = \textwidth]{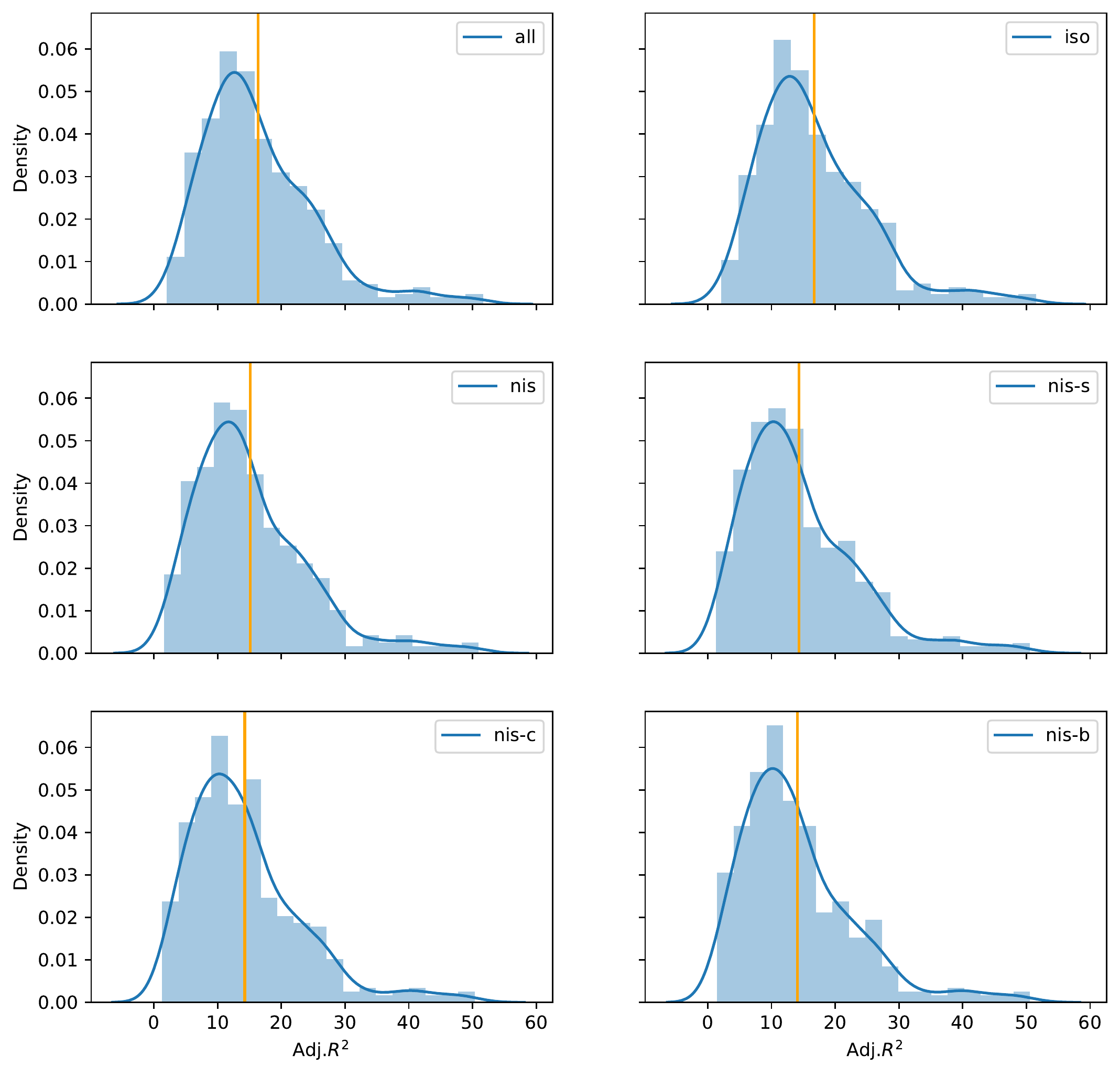}
    \caption{\updateSecondRound{Distributions of adjusted $R^2$ of contemporaneous time series regressions. \\ This table displays the histogram and kernel density estimation of the adjusted $R^2$ of contemporaneous time series regressions. The orange line indicates the mean of the adjusted $R^2$.}}
    \label{fig:contemp_R2_dist}
\end{figure}

\updateSecondRound{\Cref{tab:time_series_contemporaneous_simple_regression} summarizes the results of contemporaneous regressions. All types of COIs have positive impact on prices on average, which aligns with our findings in \Cref{contamporaneous_analysis}. \Cref{fig:contemp_coeff_dist} shows the distribution of regression coefficients. Furthermore, \Cref{fig:contemp_R2_dist} shows that the distributions of adjusted $R^2$ 
are right-skewed. }

\begin{table}[t]
    \caption{\updateSecondRound{Contemporaneous time series regressions. \\ This table summarizes the results of 457 regressions, one for each stock, using \Cref{contemporaneous_regression_equation}, against each type of COI individually. `Average $\beta_{\rho}$' denotes the mean of regressions coefficients over all stocks. `Percentage positive' denotes proportion of stocks with positive $\beta_{\rho}$. `Significant' denotes proportion of stocks with coefficients which are statistically significant at $5\%$ significance level using  a two-tailed $t$ test. `Average adj. $R^{2}$' denotes the adjusted $R^{2}$ averaged across all stocks.}} \label{tab:time_series_contemporaneous_simple_regression}
        \begin{tabularx}{\textwidth}{l * 7 {Y}}
        \toprule

                     &  Average $\beta_{\rho}$ & Standard deviation of $\beta_{\rho}$ & Percentage positive  &  Percentage significant  &  Percentage positive and significant & Average adj.$R^{2} (\%)$ \\ 
         \midrule
        all   & 2.16 & 1.20 & 98.69 & 90.81 & 90.59 & 16.36 \\
        iso   & 1.72 & 0.97 & 98.91 & 94.09 & 94.09 & 16.72 \\
        nis   & 1.56 & 1.09 & 95.62 & 81.84 & 81.18 & 15.15 \\
        nis-s & 0.95 & 1.05 & 86.21 & 63.46 & 61.71 & 14.33 \\
        nis-c & 0.93 & 0.62 & 95.62 & 76.81 & 76.59 & 14.25 \\
        nis-b & 0.77 & 0.59 & 91.90 & 71.12 & 70.46 & 14.03 \\
        \bottomrule
    \end{tabularx}

\end{table}

\subsection{Predictive Time Series Regression}

\begin{figure}[t]
    \centering
    \includegraphics[width = \textwidth]{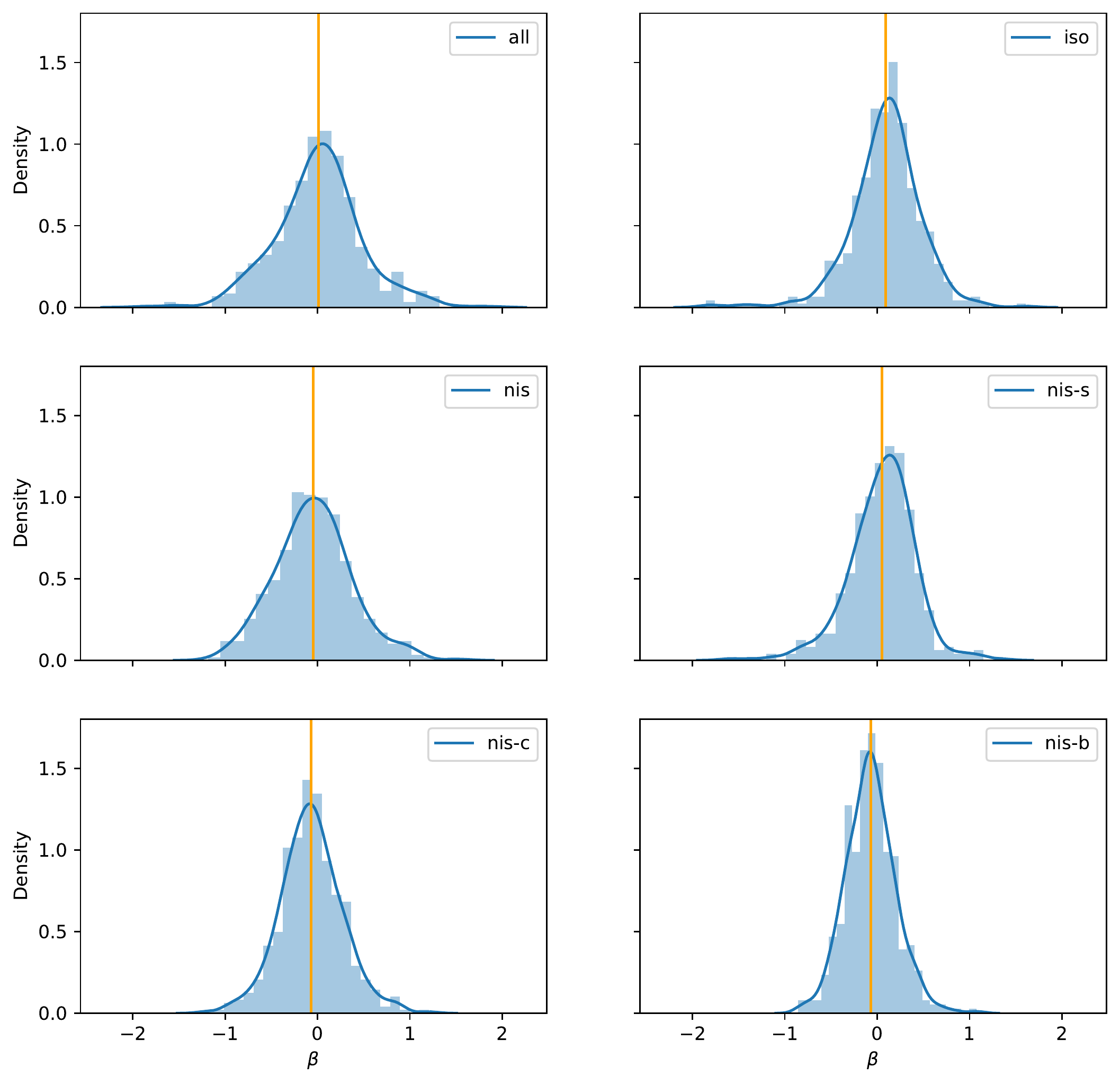}
    \caption{\updateSecondRound{Distributions of coefficients of predictive time series regressions. \\This table shows the histogram and kernel density estimation of the coefficients of predictive time series regressions. The orange line indicates the mean of the coefficients.}}
    \label{fig:pred_coeff_dist}
\end{figure}

\begin{figure}[t]
    \centering
    \includegraphics[width = \textwidth]{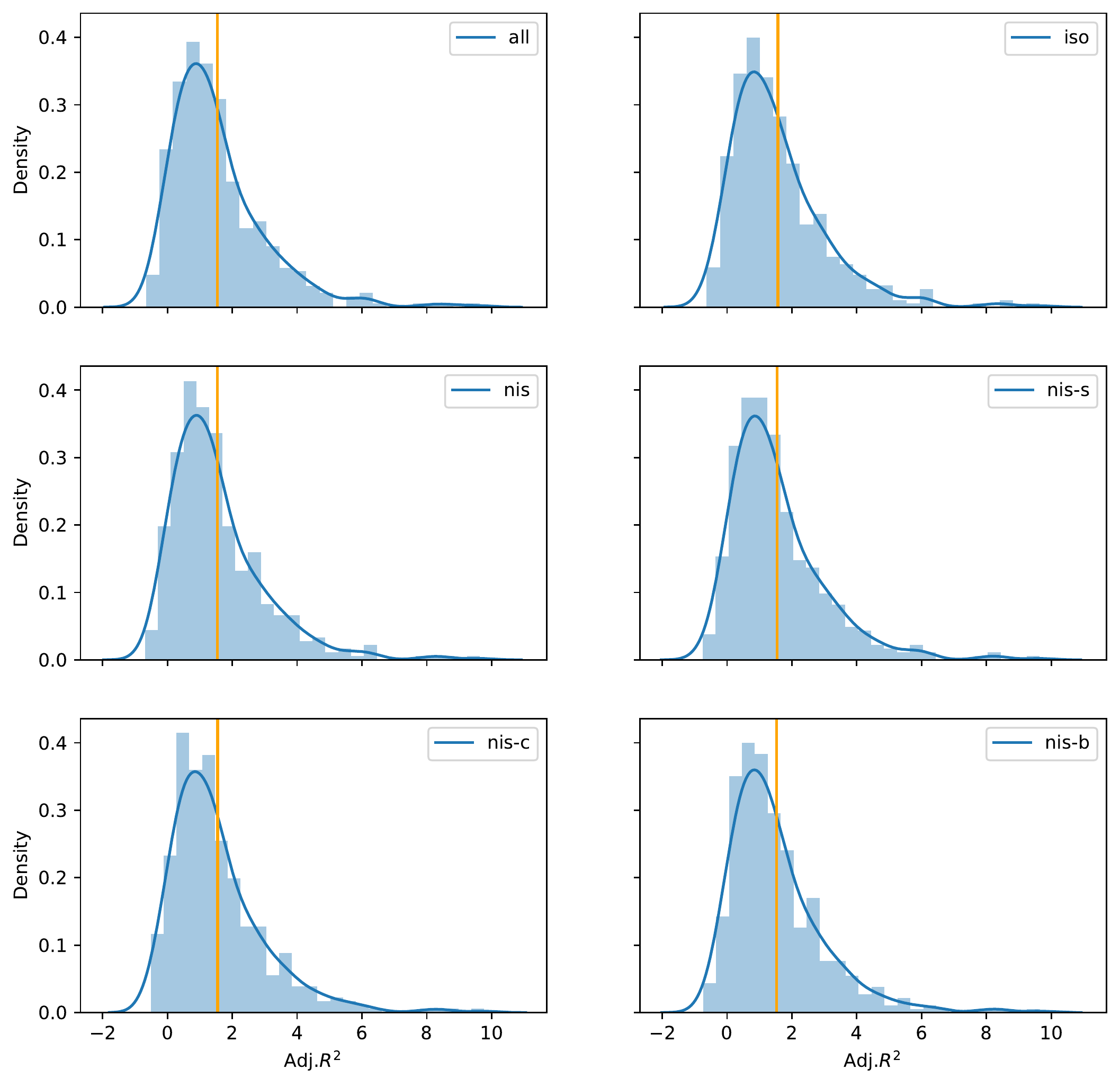}
    \caption{\updateSecondRound{Distributions of adjusted $R^2$ of predictive time series regressions. \\ This table displays the histogram and kernel density estimation of the adjusted $R^2$ of predictive time series regressions. The orange line indicates the mean of the adjusted $R^2$.}}
    \label{fig:pred_R2_dist}
\end{figure}

\updateSecondRound{\Cref{tab:time_series_predictive_simple_regression} summarizes the results of predictive regressions. The signs of the coefficients of COIs are consistent with our findings in \Cref{predictability}. \Cref{fig:pred_coeff_dist} shows the distribution of regression coefficients. Furthermore, \Cref{fig:pred_R2_dist} shows that the distributions of the adjusted $R^2$  are right-skewed. }

\begin{table}[t]
    \caption{\updateSecondRound{Predictive time series regressions. \\ This table summarizes the results of 457 regressions, one for each stock, using \Cref{predictive_regression_equation}, against each type of COI individually. `Average $\beta_{\rho}$' denotes the mean of regressions coefficients over all stocks. `Percentage positive' denotes proportion of stocks with positive $\beta_{\rho}$. `Significant' denotes proportion of stocks with coefficients which are statistically significant at $5\%$ significance level using a two-tailed $t$ test. `Average adj. $R^{2}$' denotes the adjusted $R^{2}$ averaged across all stocks.}} \label{tab:time_series_predictive_simple_regression}
        \begin{tabularx}{\textwidth}{l * 7 {Y}}
        \toprule

                     &  Average $\beta_{\rho}$ & Standard deviation of $\beta_{\rho}$ & Percentage positive  &  Percentage significant  &  Percentage positive and significant & Average adj.$R^{2} (\%)$ \\ 
         \midrule
        all   & 0.01  & 0.47 & 51.64 & 7.44 & 5.03 & 1.54 \\
        iso   & 0.09  & 0.38 & 65.86 & 9.41 & 7.88 & 1.57 \\
        nis   & -0.05 & 0.42 & 44.20 & 5.03 & 2.41 & 1.54 \\
        nis-s & 0.05  & 0.36 & 59.96 & 5.69 & 3.94 & 1.55 \\
        nis-c & -0.07 & 0.34 & 39.61 & 6.78 & 1.75 & 1.55 \\
        nis-b & -0.07 & 0.27 & 36.98 & 3.72 & 0.66 & 1.53 \\
        \bottomrule
    \end{tabularx}

\end{table}

\newpage
\section{Additional evaluation for regression analysis} \label{appendix:metrics}
\update{\Cref{tab:additional_metrics} provides additional evaluation for the regression analysis in \Cref{contamporaneous_analysis} and \Cref{predictability}. The conclusions we derive are consistent under additional evaluation. 
 }

\begin{table}[htp]
    \caption{\update{Additional evaluation for regression analysis.\\
    This table reports additional evaluation metrics, including F-score of regression, AIC, BIC, MSE and MAE. Panel A supplements the results of contemporaneous regressions in \Cref{tab:contemporaneous_simple_regression} and \Cref{tab:contemporaneous_multiple_regression}. Panel A supplements the results of predictive regressions in \Cref{tab:predictive_regression} and \Cref{tab:predictive_multiple_regression}.}} 
    \label{tab:additional_metrics}
        \begin{tabularx}{\textwidth}{l * 6 {Y}}

        \toprule
        \multicolumn{6}{l}{\textbf{Panel A: Contemporaneous regression}}  \\
        \midrule
                                        & F-score & AIC        & BIC        & MSE    & MAE    \\
        \midrule

all                                     & 3014.97 & 1632679.36 & 1632783.54 & 2.1614 & 0.9661 \\
iso                                     & 3160.07 & 1631448.98 & 1631553.16 & 2.1555 & 0.9665 \\
nis                                     & 2356.34 & 1638306.63 & 1638410.81 & 2.1884 & 0.9750 \\
nis-s                                   & 1789.82 & 1643203.51 & 1643307.69 & 2.2123 & 0.9839 \\
nis-c                                   & 2062.61 & 1640838.93 & 1640943.11 & 2.2007 & 0.9803 \\
nis-b                                   & 1824.98 & 1642898.05 & 1643002.23 & 2.2108 & 0.9827 \\
iso + nis                               & 2988.08 & 1630095.96 & 1630213.16 & 2.1491 & 0.9634 \\
iso + nis-c                             & 2977.28 & 1630197.28 & 1630314.48 & 2.1496 & 0.9639 \\
iso + nis-c + nis-b                     & 2760.67 & 1629639.74 & 1629769.96 & 2.1469 & 0.9627 \\
iso + nis-c + nis-b + nis-s             & 2550.43 & 1629416.89 & 1629560.13 & 2.1459 & 0.9625 \\
iso + nis-c + nis-b + nis-s + nis       & 2454.35 & 1628199.49 & 1628355.76 & 2.1401 & 0.9606 \\
iso + nis-c + nis-b + nis-s + all       & 2423.37 & 1628575.80 & 1628732.07 & 2.1419 & 0.9609 \\
iso + nis-c + nis-b + nis-s + nis + all & 2279.23 & 1628196.95 & 1628366.24 & 2.1401 & 0.9606 \\   
        \bottomrule
    \end{tabularx}        
    
    \begin{tabularx}{\textwidth}{l * 6 {Y}}

        \multicolumn{6}{l}{\textbf{Panel B: Predictive regression}}  \\
        \midrule
                                        & F-score & AIC        & BIC        & MSE    & MAE    \\
        \midrule
all                                     & 46.96         & 1655230.52 & 1655347.72 & 2.2719 & 0.9950 \\
iso                                     & 50.47         & 1655195.40 & 1655312.66 & 2.2717 & 0.9949 \\
nis                                     & 47.77        & 1655222.42 & 1655339.62 & 2.2718 & 0.9950 \\
nis-s                                   & 47.63         & 1655223.81 & 1655341.02 & 2.2718 & 0.9950 \\
nis-c                                   & 48.60        & 1655214.09 & 1655331.29 & 2.2718 & 0.9949 \\
nis-b                                   & 50.65         & 1655193.67 & 1655310.87 & 2.2717 & 0.9949 \\
iso + nis                               & 49.77         & 1655152.70 & 1655282.93 & 2.2715 & 0.9948 \\
iso + nis-c                             & 49.87        & 1655151.64 & 1655281.87 & 2.2715 & 0.9948 \\
iso + nis-c + nis-b                     & 47.55        & 1655129.59 & 1655272.83 & 2.2714 & 0.9948 \\
iso + nis-c + nis-b + nis-s             & 43.94        & 1655128.98 & 1655285.25 & 2.2714 & 0.9948 \\
iso + nis-c + nis-b + nis-s + nis       & 40.86        & 1655128.16 & 1655297.45 & 2.2713 & 0.9948 \\
iso + nis-c + nis-b + nis-s + all       & 41.01        & 1655126.12 & 1655295.41 & 2.2713 & 0.9948 \\
iso + nis-c + nis-b + nis-s + nis + all & 38.34        & 1655125.05 & 1655307.36 & 2.2713 & 0.9948 \\
        \bottomrule
    \end{tabularx}    
\end{table}

\newpage
\section{Neighbourhood size effect} \label{appendix:neighbourhood_size}

\begin{figure}[t]
    \includegraphics[width = \textwidth]{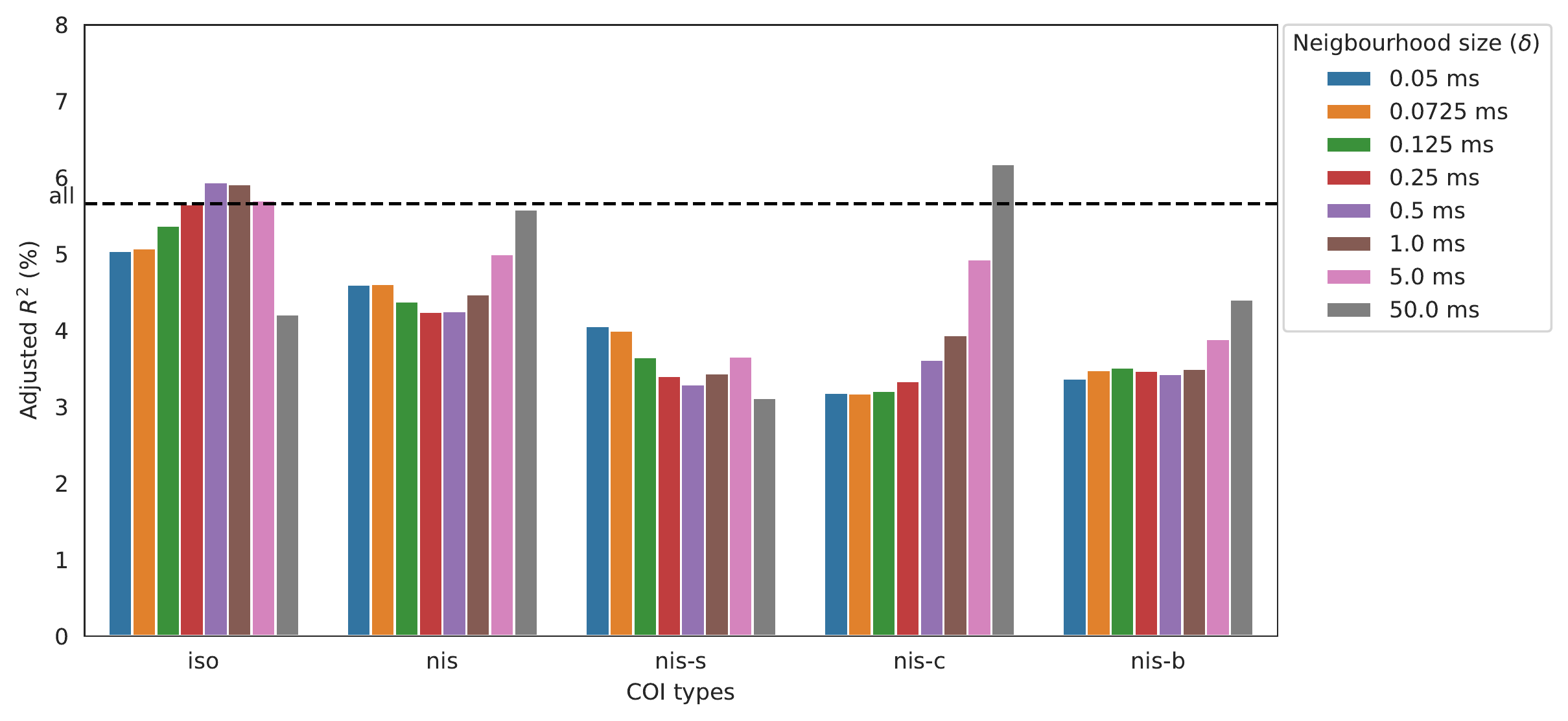}
    \caption{Average $R^2$ of regressing contemporaneous returns against COIs for different $\delta$s.}
    \label{fig:contemporaneous_regression_by_delta}
\end{figure}
    
\begin{figure}[h]
    \includegraphics[width = \textwidth]{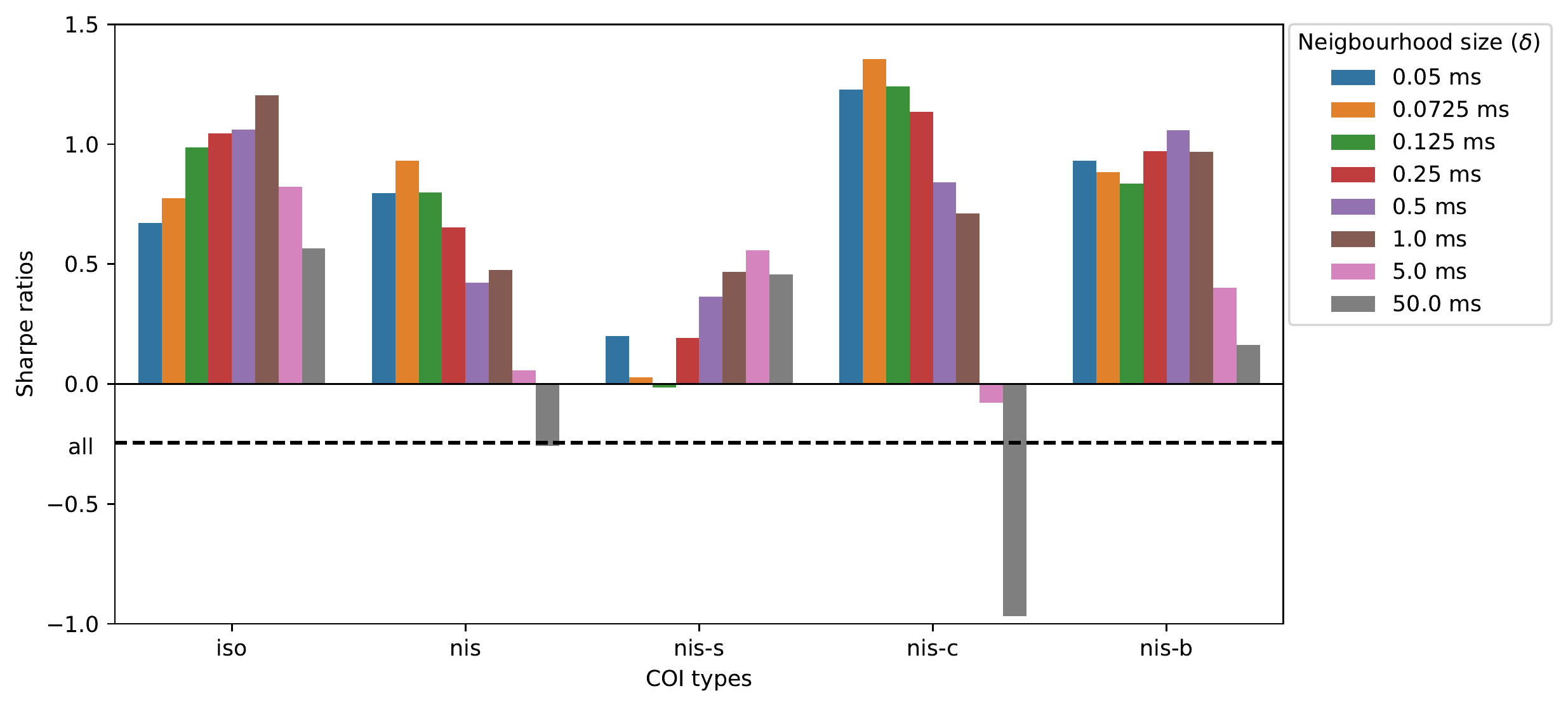}
    \caption{Sharpe ratios of COI-based long-short portfolios for different $\delta$s.} 
    \label{fig:sharpe_by_delta}
\end{figure}
    

To study the effect of neighbourhood size on conditional order imbalances, we repeat the regression and portfolio analysis for each $\delta \in \{$0.05 ms, 0.075 ms, 0.125 ms, 0.25 ms, 0.5 ms, 1 ms, 5 ms, 50 ms$\}$, and display the results.

\Cref{fig:contemporaneous_regression_by_delta} illustrates the average $R^2$ of contemporaneous regressions. Isolated order imbalances achieve the highest $R^2$ at $\delta = 0.5$ ms. In contrast, the histograms of $R^2$ of the  non-isolated imbalances have a U-shape with minimum  at $\delta = 0.5$ ms. For the three types of non-isolated order imbalances, the $R^2$s for non-self-isolated and non-both-isolated imbalances have downward trends with growth in values of $\delta$. Non-cross-isolated imbalances explain more variance in returns as $\delta$ increases. 

\Cref{fig:sharpe_by_delta} details the Sharpe Ratios of long-short portfolios of different COI types ordered by $\delta$. We remark that the Sharpe Ratios of each type of order imbalance peak at different values of $\delta$.

\newpage
\section{Representative of the market effect} \label{appendix:universe_stock}

\updateSecondRound{The classification of trades depends on the set of stocks we choose as the market index, $\mathcal{M}$. In this section, we compare the fractions of trades, COIs and economic values of the same 457 stocks as described in \Cref{data}, while using constituents of S\&P 500, S\&P 100, and Dow 30 indices as $\mathcal{M}$, respectively, for the trade flow decomposition. Our original universe contains 457 S\&P 500 companies, and we decompose trade flows for all of them, based on the intersections with the other two, smaller, indices. \Cref{tab:universe_corr} reports the results.}   


\begin{table}[t]
    \caption{\updateSecondRound{Trades and COIs by universe of stocks. \\ This table shows the results 
    for three sets of stocks as the market index, including S\&P 500, S\&P 100 and Dow 30. Panel A documents the average fractions of each type of trade for each universe of stocks. Panel B reports the correlations of all types of COIs for each pair of universes. The columns indicate the pairs. Panel C presents the annualized Sharpe ratios, given by \Cref{sharpe_ratio_equation}, of single-sort and selected double-sort long-short portfolios based on COIs of the universes of stocks.}}    
        \begin{tabularx}{\textwidth}{l*{4}{Y}}
        \toprule
        \multicolumn{4}{l}{\textbf{Panel A: Average fractions}}          \\
        \midrule
                  &  S\&P 500      &  S\&P 100     &  Dow 30      \\ 
        \cmidrule(lr){2-4}
                iso   &     28.55 &     34.37 &   37.57 \\
                nis   &     71.45 &     65.63 &   62.43 \\
                nis-s &     29.75 &     37.82 &   42.58 \\
                nis-c &     17.27 &     11.45 &    8.24 \\
                nis-b &     24.43 &     16.37 &   11.60 \\
    \end{tabularx}
    
    \begin{tabularx}{\textwidth}{l*{4}{Y}}
        \toprule
        \multicolumn{4}{l}{\textbf{Panel B: Average correlations}}          \\
        \midrule
                 &  S\&P 500 - S\&P 100  &  S\&P 500 - Dow 30   &  S\&P 100 - Dow 30 \\ 
         \cmidrule(lr){2-4}
        iso   &               0.99 &              0.97 &              0.99 \\
        nis   &               0.99 &              0.99 &              1.00 \\
        nis-s &               0.97 &              0.94 &              0.98 \\
        nis-c &               0.92 &              0.85 &              0.95 \\
        nis-b &               0.91 &              0.82 &              0.91 \\
    \end{tabularx}
    
    \begin{tabularx}{\textwidth}{l*{4}{Y}}
        \toprule
        \multicolumn{4}{l}{\textbf{Panel C: Annualized Sharpe ratios}}          \\
        \midrule
                  &  S\&P 500      &  S\&P 100     &  Dow 30      \\ 
         \cmidrule(lr){2-4}
        iso       &     1.20 &     1.04 &    0.84 \\
        nis       &     0.48 &     0.59 &   -0.98 \\
        nis-s     &     0.47 &     0.26 &    0.02 \\
        nis-c     &     0.71 &     1.12 &   -1.59 \\
        nis-b     &     0.97 &     1.16 &   -1.65 \\
        iso/nis   &     1.08 &     1.22 &    0.76 \\
        iso/nis-c &     1.79 &     1.34 &    1.03 \\
        \bottomrule
    \end{tabularx}
    \label{tab:universe_corr}
\end{table}

\newpage
\section{Time-of-day effect}  \label{appendix:time_effect}

\begin{table}[t]
    \caption{COIs by time period. \\ We calculate COIs of 9:30 $-$ 10:00, 10:00 $-$ 15:30, and 15:30 $-$ 16:00 each day from 2017-01-03 to \update{2020-12-31} for the selected 457 stocks. Panel A summarizes the adjusted $R^2$ of regressions on contemporaneous open-to-close market excess returns against each type of COIs, using \Cref{contemporaneous_regression_equation}. Panel B presents the annualized Sharpe Ratios, given by \Cref{sharpe_ratio_equation}, of single-sort long-short portfolios based on COIs of different intraday time periods. The last column of both panels reports the daily COIs as a benchmark.} 
        \begin{tabularx}{\textwidth}{l * 5 {Y}}

        \toprule
        \multicolumn{5}{l}{\textbf{Panel A: Contemporaneous regression $R^2 (\%)$ }}          \\
        \midrule
                     &  9:30 $-$ 10:00  &  10:00 $-$ 15:30  &  15:30 $-$ 16:00 &  9:00 $-$ 16:00 \\ 
                     \cmidrule(lr){2-5}  
         all         &   3.64          &   5.16            &   3.75        &  5.66   \\
         iso         &   4.70          &   5.32            &   3.18        &  5.91   \\
         nis         &   2.93          &   4.17            &   3.56        &  4.48   \\
         nis-s       &   2.89          &   3.27            &   2.84        &  3.44  \\
         nis-c       &   3.04          &   3.69            &   3.00        &   3.94  \\
         nis-b       &   2.65          &   3.36            &   3.09        &   3.50  \\
        \bottomrule
    \end{tabularx}

    \begin{tabularx}{\textwidth}{l * 5 {Y}}
        \multicolumn{5}{l}{\textbf{Panel B: Annualized Sharpe ratios}}          \\
        \midrule
                     &  9:30 $-$ 10:00  &  10:00 $-$ 15:30  &  15:30 $-$ 16:00 &  9:00 $-$ 16:00 \\ 
                     \cmidrule(lr){2-5}  
         all         & -1.43            &  -0.67           &   0.43          &  -0.25  \\
         iso         &  -0.94           &  1.33            &   1.61          &  1.20   \\
         nis         &  0.82           &  0.40            &   -0.56         &   0.48  \\
         nis-s       &  -0.85           &  0.18            &   1.06          &  0.47   \\
         nis-c       &  1.03           &  0.64            &   -0.53         &   0.71  \\
         nis-b       &  0.74           &  0.83            &   0.20          &   0.97  \\
        \bottomrule
    \end{tabularx}
    \label{tab:coi_by_time}
\end{table}

We investigate the COIs of different intraday time intervals. Firstly, we evaluate their influences on same-day price change by regressing contemporaneous open-to-close market access returns against each COI individually. Panel A of \Cref{tab:coi_by_time} presents the  $R^{2}$ of all such regressions. Excluding the first and last half hours of trades does not explicitly change the imbalance-return relations we discover. Regardless of periods, deriving COIs with only \textit{iso} trades is enough to explain a comparable amount of variance as when using all trades. Note that, especially for the first hour, the price impact mainly stems from isolated trades.

Secondly, we trade on each COI by constructing single-sort long-short portfolios and present annualized Sharpe Ratios in Panel B. It is reasonable to expect that trading activities towards the end of the normal trading period contribute more to forecasting future returns. We observe that the signal corresponding to the \textit{iso} and \textit{nis-s} COIs of the last hour leads to a 0.41 and 0.59 increase in Sharpe Ratios, significantly enhancing the portfolio profits. Conversely, the last half-hour of non-cross-isolated COI is not a good signal for predicting future returns. For the  \textit{nis-b} trades, the future returns are only predicted by the  COI during less volatile trading hours.

\newpage
\section{COI measured by volumes} \label{appendix:volume_coi}
Instead of considering the number of trades, in this section we analyze volume order imbalances defined as

\begin{equation} \label{coi_volume_equation}
    COI_{i, t}^{type} = \frac{V_{i, t}^{type, buy} - V_{i, t}^{type, sell}}{V_{i, t}^{type, buy} + V_{i, t}^{type, sell}},
\end{equation}
where $V_{i, t}^{type, buy}$ and $V_{i, t}^{type, sell}$ denote the total volume of market buy orders and market sell orders of stock $i$ on day $t$. We repeat the analysis on volume imbalances and present the results in \Cref{tab:coi_by_volume}.  

\begin{table}[ht]
    \caption{COIs measured by volume. \\ We calculate COIs measured by volumes, as \autoref{coi_volume_equation}, from 2017-01-03 to \update{2020-12-31} for the selected 457 stocks. Panel A summarizes the coefficients to COIs by regressing again each type of COIs individually following \Cref{contemporaneous_regression_equation}. `$\beta_{\rho}$' denotes the regression coefficients and the superscript $\text{***}$ indicates significant at $1\%$ using two-tailed t-test. `$t$' denotes the t-value of each coefficient. `adj.$R^2$' denotes the adjusted $R^{2}$ of regressions. Panel B shows the annualized Sharpe Ratios of the long-short portfolios sorted on COIs indicated by the corresponding row indices and column names. The on- and off- diagonal values are for single- and double-sort portfolios respectively. The annualized Sharpe Ratios over the sample period are given by \autoref{sharpe_ratio_equation}. }    
        \begin{tabularx}{\textwidth}{l * 4 {Y}}
        \toprule
        \multicolumn{4}{l}{\textbf{Panel A: Contemporaneous regression}}          \\
        \midrule
        
                              &  $\beta_{\rho}$ & t   &  adj.$R^2$ (\%)   \\ 
        \midrule  
        
            all   & 1.58*** & 24.72 & 4.70 \\
            iso   & 1.30*** & 30.85 & 5.02 \\
            nis   & 1.19*** & 19.21 & 3.90 \\
            nis-s & 0.75*** & 15.52 & 3.37 \\
            nis-c & 0.58*** & 16.24 & 3.14 \\
            nis-b & 0.61*** & 16.46 & 3.27 \\
        
    \end{tabularx}    
    
    \begin{tabularx}{\textwidth}{l*{6}{Y}}
        \toprule
        \multicolumn{6}{l}{\textbf{Panel B: Annualized Sharpe ratios}}          \\
        \midrule
                 &  iso  &  nis   &  nis-s  &  nis-c  &  nis-b \\ 
        \midrule
        iso      & 0.88  &  1.23  &  0.15   &  1.49   &  0.93   \\
        nis      &       &  0.59  &  -0.29  &  0.76   &  0.72   \\
        nis-s    &       &        &  0.24   &  -0.30  &  -0.19  \\
        nis-c    &       &        &         &   0.91  &  0.86   \\
        nis-b    &       &        &         &         &  1.01       \\
        \bottomrule
    \end{tabularx}
    \label{tab:coi_by_volume}
\end{table}

\newpage
\section{Further analysis on portfolio profitability} \label{appendix:portfolio_profit}

\updateSecondRound{To verify the robustness of the profitability of the proposed COI-sorted portfolios, we apply transaction costs on backtests. Assuming flat round trip transaction costs over all stocks, we test on cost rates including 1, 2, 3, 4 and 5 bps. Recall that, for the sort-based long-short strategies, we open positions at market open and liquidate at market, without holding overnight positions; the daily turnover is always 100\%. Therefore, we directly subtract fixed transaction cost rates from daily portfolio returns during backtesting. Additionally, we ignore transaction costs for equal weight and SPY ETF since daily rebalancing is not needed for them. \Cref{tab:port_with_cost} reports the annualized returns and Sharpe ratios selected portfolios and benchmarks. From the table, we observe that the portfolio single-sorted on \textit{iso} turns to loss when cost is greater than 2 bps. 
In contrast, the profitability of portfolios double-sorted on \textit{iso$-$nis} and \textit{iso$-$nis-c} persists and consistently outperform benchmarks. In particular, the \textit{iso$-$nis-c} portfolio obtains annualized return of 22.27\% and Sharpe ratio of 1.11 under the strictest scenario.}

\begin{table}[t]
    \caption[LoF entry]{\updateSecondRound{Annualized returns and Sharpe ratios of selected and benchmark portfolios. \\ This table exhibits the results of selected and benchmark portfolios in \Cref{fig:cum_return}. The column names indicate different levels of transaction costs in basis points (bps). Panel A presents the annualized return of portfolios calculated by averaging their daily returns, from 2017-01-03 to 2020-12-31, and multiplying by 252. Panel B reports the annualized Sharpe ratios over the sample period calculated by \Cref{sharpe_ratio_equation}.}} 
    \label{tab:port_with_cost}
    
    \begin{tabularx}{\textwidth}{l*{7}{Y}}
\toprule
\multicolumn{7}{l}{\textbf{Panel A: Annualized returns}} \\
\midrule
portfolio / cost (bps) & 0     & 1     & 2     & 3     & 4     & 5     \\
\midrule
iso                     & 6.76   & 4.24   & 1.72   & -0.80  & -3.32  & -5.84  \\
iso/nis                 & 23.33  & 20.81  & 18.29  & 15.77  & 13.25  & 10.73  \\
iso/nis-c               & 34.87  & 32.35  & 29.83  & 27.31  & 24.79  & 22.27  \\
all                     & 0.31   & -2.21  & -4.73  & -7.25  & -9.77  & -12.29 \\
return momentum         & -14.81 & -17.33 & -19.85 & -22.37 & -24.89 & -27.41 \\
equal weight            & -4.00  &  &  &  &  &      \\
SPY ETF (open to close) & -1.00  & -3.52  & -6.04  & -8.56  & -11.08 & -13.6  \\
SPY ETF                 & 10.04  &  &  &  &  &  \\
\end{tabularx}

\begin{tabularx}{\textwidth}{l*{7}{Y}}
\toprule
\multicolumn{7}{l}{\textbf{Panel B: Annualized Sharpe ratios}} \\
\midrule
portfolio / cost (bps) & 0     & 1     & 2     & 3     & 4     & 5     \\
\midrule
iso                                 & 1.20  & 0.62  & 0.03  & -0.55 & -1.14 & -1.72 \\
iso/nis                             & 1.08  & 0.95  & 0.83  & 0.70  & 0.58  & 0.45  \\
iso/nis-c                           & 1.79  & 1.66  & 1.52  & 1.39  & 1.25  & 1.11  \\
all                                 & -0.25 & -0.74 & -1.23 & -1.71 & -2.20 & -2.69 \\
return momentum                     & -1.23 & -1.41 & -1.60 & -1.79 & -1.98 & -2.17 \\
equal weight                        & -0.39 &  &  &  &  &  \\
SPY ETF (open to close)             & -0.19 & -0.38 & -0.57 & -0.76 & -0.95 & -1.14 \\
SPY ETF                             & 0.42 &  &  &  &  &  \\
\bottomrule
\end{tabularx}

\end{table}

\end{document}